\documentclass[showpacs,aps,prd,preprint,nofootinbib,showkeys,unsortedaddress,raggedbottom]{revtex4-1}
\pdfoutput=1

\usepackage{bm}
\usepackage{amsmath}
\usepackage{graphicx}
\usepackage{subfigure}
\usepackage[usenames,dvipsnames]{color}
\definecolor{darkblue}{RGB}{0,0,196}
\usepackage[colorlinks=true,linkcolor=darkblue,citecolor=darkblue,urlcolor=darkblue]{hyperref}

\usepackage{setspace}
\usepackage{footmisc}
\usepackage[makeroom]{cancel}

\newcommand{\intdP}{\int\!dP}

\def\be{\begin{equation}}
\def\ee{\end{equation}}
\def\ba{\begin{eqnarray}}
\def\ea{\end{eqnarray}}

\begin{document}

\title{Quasiparticle equation of state for anisotropic hydrodynamics}

\author{Mubarak Alqahtani} 

\author{Mohammad Nopoush} 

\author{Michael Strickland} 
\affiliation{Department of Physics, Kent State University, Kent, OH 44242 United States}

\begin{abstract}
We present a new method for imposing a realistic equation of state in anisotropic hydrodynamics.  The method relies on the introduction of a single finite-temperature quasiparticle mass which is fit to lattice data.  By taking moments of the Boltzmann equation, we obtain a set of coupled partial differential equations which can be used to describe the 3+1d spacetime evolution of an anisotropic relativistic system.  We then specialize to the case of a 0+1d system undergoing boost-invariant Bjorken expansion and subject to the relaxation-time approximation collisional kernel.  Using this setup, we compare results obtained using the new quasiparticle equation of state method with those obtained using the standard method for imposing the equation of state in anisotropic hydrodynamics.  We demonstrate that the temperature evolution obtained using the two methods is nearly identical and that there are only small differences in the pressure anisotropy.  However, we find that there are significant differences in the evolution of the bulk pressure correction.
\end{abstract}

\date{\today}



\maketitle

\section{Introduction}
\label{sec:intro}

Ultrarelativistic heavy-ion collision experiments using the Relativistic Heavy Ion Collider (RHIC) at Brookhaven National Laboratory and the Large Hadron Collider (LHC) at CERN allow researchers to study the behavior of matter subject to extreme conditions.  In these experiments, high-energy collisions of nuclei are used to heat a tiny volume of matter up to temperatures that exceed the critical temperature ($T_c\sim160 \; \rm MeV$) necessary to create a super-hot deconfined and chirally-symmetric phase, called the quark-gluon plasma (QGP).  The study of this strongly interacting phase near and above the critical temperature is of fundamental interest.  One can gain some insight into the physics of the QGP using perturbation theory since the asymptotic freedom of quantum chromodynamics (QCD) ensures that, for the high temperatures, $T \gg \Lambda_{\rm QCD}$, the QGP can be thought of as a weakly-coupled many-body system. In this regime, perturbative methods, such as hard thermal loop (HTL) resummation, can be used \cite{Braaten:1989mz,Andersen:1999fw,Andersen:2003zk,Andersen:2004fp,Andersen:2011sf,Haque:2014rua}.\footnote{HTL-resummed calculations of the thermodynamic potential at finite temperature and quark chemical potential(s) describe the lattice data well for $T \gtrsim 300$ MeV with no free parameters~\cite{Haque:2014rua,Bellwied:2015lba,Ding:2015fca}.}  In the HTL framework, the quarks and gluons can be thought of as quasiparticles having temperature-dependent (thermal) masses with $m_{q,\bar{q},g} \sim g T$, where $g$ is the strong coupling.

Such a picture provides motivation to try to model the QGP as a gas of massive quasiparticles for the purposes of obtaining self-consistent hydrodynamic equations.  However, perturbation theory needs to be supplemented since, for temperatures $T \lesssim 2 T_c$, first-principles perturbative calculations based on deconfined quarks and gluons break down.  In order to proceed, one can use non-perturbative lattice calculations of QCD thermodynamics to determine information about the necessary quasiparticle mass(es).  In practice, one can perform this procedure at all temperatures and determine a non-perturbative temperature-dependent quasiparticle mass, $m(T)$.  Once $m(T)$ is determined, one can use this to enforce the target equation of state (EoS) in an effective kinetic field theory framework.  One complication is that, in order to guarantee thermodynamic consistency in equilibrium and related out-of-equilibrium constraints, it is necessary to introduce a background (vacuum energy) contribution to the energy-momentum tensor~\cite{gorenstein1995gluon,Jeon:1995zm,Romatschke:2011qp}.  The resulting EoS, together with a self-consistent non-equilibrium energy-momentum tensor and modified Boltzmann equation, can be used to derive relativistic hydrodynamic equations for such a quasiparticle gas.

Relativistic hydrodynamics itself is an effective theory that can be used to describe the spacetime evolution of the QGP. In the kinetic theory approach to relativistic hydrodynamics, one obtains the dynamical equations for the bulk variables by taking moments of the Boltzmann equation. Ideal hydrodynamics \cite{Huovinen:2001cy,Hirano:2002ds,Kolb:2003dz} and later on viscous hydrodynamics \cite{Muronga:2001zk,Muronga:2003ta,Muronga:2004sf,Heinz:2005bw,Baier:2006um,Romatschke:2007mq,Baier:2007ix,Dusling:2007gi,Luzum:2008cw,Song:2008hj,Heinz:2009xj,Bozek:2009ty,Bozek:2009dw,El:2009vj,PeraltaRamos:2009kg,PeraltaRamos:2010je,Denicol:2010tr,Denicol:2010xn,Schenke:2010rr,Schenke:2011tv,Bozek:2011wa,Bozek:2011ua,Niemi:2011ix,Niemi:2012ry,Bozek:2012qs,Denicol:2012cn,Denicol:2012es,PeraltaRamos:2012xk,Calzetta:2014hra,Denicol:2014vaa,Florkowski:2015lra,Ryu:2015vwa} have been used to study the QGP created in heavy-ion collisions and have proven to be quite successful. Recently, anisotropic hydrodynamics \cite{Martinez:2010sc,Florkowski:2010cf,Ryblewski:2010bs,Martinez:2010sd,Ryblewski:2011aq,Florkowski:2011jg,Martinez:2012tu,Ryblewski:2012rr,Florkowski:2012as,Florkowski:2013uqa,Ryblewski:2013jsa,Bazow:2013ifa,Tinti:2013vba,Florkowski:2014bba,Florkowski:2014txa,Nopoush:2014pfa,Denicol:2014mca,Nopoush:2014pfa,Tinti:2015xra,Bazow:2015cha,Nopoush:2015yga,Bazow:2015zca,Florkowski:2015cba} has been developed in order to extend the range of applicability of relativistic hydrodynamics to situations in which the QGP possess a high degree of momentum-space anisotropy (for a recent review, see Ref.~\cite{Strickland:2014pga}). 

In most cases, however, the manner in which the EoS is imposed in hydrodynamics is somewhat uncontrolled.  In many cases, one derives the hydrodynamic equations for a conformal system and then imposes an EoS to relate the components of the energy-momentum tensor.  We refer to this as the ``standard EoS'' method.  However, since QCD is a non-conformal theory with a running coupling constant that depends strongly on the temperature near $T_c$, it is more self-consistent to take into account the breaking of conformal invariance from the beginning, which results in additional terms in the evolution equations and new transport coefficients.  Some progress in this direction has been made in the last year, both in the context of second-order viscous hydrodynamics \cite{Denicol:2014vaa} and anisotropic hydrodynamics \cite{Nopoush:2014pfa}, however, in both of these previous works, the underlying microscopic picture was that of a gas of particles with temperature-independent masses.  One would like to incorporate the temperature-dependence of the particle masses into the dynamical equations such that the equations themselves are consistent with the breaking of conformal invariance and the quasiparticle picture at high temperatures.  In this paper, we present a method for doing this in the context of anisotropic hydrodynamics.  Our method is to incorporate the effects of a temperature-dependent quasiparticle mass into the Boltzmann equation by taking into account extra terms which come from the spacetime gradients of the thermal mass.  We show that adding the necessary additional term to the Boltzmann equation and enforcing energy-momentum conservation require one to introduce a non-equilibrium background field $B g^{\mu\nu}$ to the energy-momentum tensor as was found by previous authors~\cite{Jeon:1995zm,Romatschke:2011qp}, e.g.
\be
T^{\mu\nu} = T^{\mu\nu}_{\rm kinetic} + B g^{\mu\nu} \, .
\ee
This extra background contribution can be shown to become precisely the additional term necessary to enforce thermodynamic consistency in the equilibrium limit, however, in practice, we allow it to be a non-equilibrium quantity.

The new method above will be referred to herein as the ``quasiparticle EoS''.  We compare results obtained using this method to results obtained using the canonical method for imposing a realistic equation of state.  For this purpose, we reduce the dynamical equations in both cases to those appropriate for 0+1d boost-invariant and transversally-homogeneous expansion subject to a relaxation-time approximation collisional kernel.  With this setup, comparisons of the evolution of the effective temperature, pressure anisotropy, and bulk correction to the pressure for different values of the shear viscosity to entropy density ratio are presented for both isotropic and anisotropic initial conditions.  We demonstrate that the temperature evolution obtained using the two EoS methods is nearly identical and that there are only small differences in the pressure anisotropy.  However, we find that there are significant differences in the evolution of the bulk pressure correction, which could potentially be important for determining the correct form of the particle distribution function on the freezeout hypersurface in phenomenological applications.

The structure of the paper is as follows. In Sec.~\ref{sec:conventions}, we present the notation and conventions we use in the paper.  In Sec.~\ref{sec:setup}, we review the necessary setup including the anisotropic distribution function, basis vectors necessary in different cases, and the lattice-based equation of state we will use.  In Sec.~\ref{sec:boltzman}, the Boltzmann equation and its generalization to quasiparticles with temperature-dependent masses is discussed. In Sec.~\ref{sec:bulk-var}, we take different moments of distribution function in order to derive expressions for the particle current, energy density, and components of the pressure.  In Sec.~\ref{sec:dynamical-eqs}, the 3+1d dynamical equations for massive anisotropic hydrodynamics are derived and then simplified assuming boost-invariance together with either cylindrical-symmetry or transversally-homogeneity. In Sec.~\ref{sec:0p1}, we obtain the 0+1d dynamical equations for the quasiparticle EoS and standard EoS cases.  In Sec.~\ref{sec:results}, our numerical results obtained using both methods for a boost-invariant and transversally-homogeneous system are presented. Sec.~\ref{sec:conclusions} contains our conclusions and an outlook for the future.  All necessary identities and function definitions are collected in Apps.~\ref{app:identities}-\ref{app:h-functions}.  

\section{Conventions and Notation}
\label{sec:conventions}

A parentheses in the indices indicates a symmetrized form, e.g. $A^{(\mu\nu)} \equiv (A^{\mu\nu} + A^{\nu\mu})/2$.  The metric is taken to be ``mostly minus'' such that in Minkowski space with $x^\mu\equiv(t,x,y,z)$ the line element is $ds^2=g_{\mu\nu} dx^\mu dx^\nu=dt^2-dx^2-dy^2-dz^2$.  We also make use of the transverse projector, $\Delta^{\mu\nu} \equiv g^{\mu\nu} - u^\mu u^\nu$.  When studying relativistic heavy-ion collisions, it is convenient to transform to variables defined by $\tau =\sqrt{t^2 - z^2}$, which is the longitudinal proper time, and $\varsigma ={\rm tanh}^{-1}(z/t)$, which is the longitudinal spacetime rapidity.   If the system is additionally cylindrically symmetric with respect to the beam-line, it is convenient to transform to polar coordinates in the transverse plane with $r=\sqrt{x^{2}{+}y^{2}}$ and $\phi ={\rm tan}^{-1}(y/x)$.   In this case, the new set of coordinates $x^\mu=(\tau,r,\phi,\varsigma)$ defines polar Milne coordinates.  Finally, the invariant phase space integration measure is defined as
\be
dP \equiv N_{\rm dof} \frac{d^3p}{(2\pi)^3} \frac{1}{E} = \tilde{N} \frac{d^3p}{E} \, ,
\ee
where $N_{\rm dof}$ is the number of degrees of freedom and $\tilde{N} \equiv N_{\rm dof}/(2\pi)^3$.

\section{Setup}
\label{sec:setup}

In this paper, we derive non-conformal anisotropic hydrodynamics equations for a system of quasiparticles with a temperature-dependent mass. To accomplish this goal, an effective Boltzmann equation for thermal quasiparticles is obtained.  We then take moments of the resulting kinetic equation to obtain the leading-order 3+1d anisotropic hydrodynamics equations.   Using a general set of basis vectors, the equations are expanded explicitly and then various simplifying assumptions (e.g. boost-invariance, etc.) are imposed to reduce the equations from their general 3+1d to a 0+1d form appropriate for describing a boost-invariant and transversally-homogenous QGP.  The obtained 0+1d equations are then solved numerically for our tests, however, the method used to obtain the 3+1d leading-order anisotropic hydrodynamics equations can be used without lack of generality.

\subsection{Basis Vectors}
\label{subsec:basis}

A general tensor basis can be constructed by introducing four four-vectors which in the local rest frame (LRF) are 
\ba
u^\mu_{\rm LRF}  &\equiv & (1,0,0,0)\,,\nonumber \\
X^\mu_{\rm LRF}  &\equiv & (0,1,0,0)\,,\nonumber \\
Y^\mu_{\rm LRF}  &\equiv & (0,0,1,0)\,,\nonumber \\
Z^\mu_{\rm LRF}  &\equiv & (0,0,0,1)\,.
\ea

One can define the general basis vectors in the lab frame (LF) by performing the Lorentz transformation necessary to go from LRF to the LF.  The transformation required can be constructed using a longitudinal boost $\vartheta$ along the beam axis, followed by a rotation $\varphi$ around the beam axis, and finally a transverse boost by $\theta_\perp$ along the $x$-axis~\cite{Florkowski:2011jg,Martinez:2012tu}. This parametrization gives
\ba
u^\mu &\equiv& \left(\cosh\theta_\perp \cosh\vartheta,\sinh\theta_\perp\cos\varphi,\sinh\theta_\perp\sin\varphi,\cosh\theta_\perp \sinh\vartheta\right) ,  \nonumber \\
X^\mu &\equiv& \left(\sinh\theta_\perp \cosh\vartheta,\cosh\theta_\perp\cos\varphi,\cosh\theta_\perp\sin\varphi,\sinh\theta_\perp \sinh\vartheta\right) , \nonumber  \\ 
Y^\mu &\equiv& \left(0,-\sin\varphi,\cos\varphi,0\right) , \nonumber  \\
Z^\mu &\equiv& \left(\sinh\vartheta,0,0,\cosh\vartheta\right) ,
\label{eq:basisvecs}
\ea
where the three fields $\vartheta$, $\varphi$, and $\theta_\perp$ are functions of Cartesian Milne coordinates $(\tau,x,y,\varsigma)$. Introducing another parametrization by using the temporal and transverse components of flow velocity
\ba 
u_0&=&\cosh\theta_\perp\, , \\
u_x&=& u_\perp \cos\varphi\, , \\
u_y&=& u_\perp \sin\varphi\, , 
\label{eq:u-par}
\ea
where $u_\perp\equiv \sqrt{u_x^2+u_y^2}=\sqrt{u_0^2-1} = \sinh\theta_\perp$, one has
\ba
u^\mu &\equiv& (u_0 \cosh\vartheta,u_x,u_y,u_0 \sinh\vartheta) \, , \nonumber\\
X^\mu &\equiv& \Big(u_\perp\cosh\vartheta,\frac{u_0 u_x}{u_\perp},\frac{u_0 u_y}{u_\perp},u_\perp\sinh\vartheta\Big) , \nonumber \\ 
Y^\mu &\equiv& \Big(0,-\frac{u_y}{u_\perp},\frac{u_x}{u_\perp},0\Big)  , \nonumber \\
Z^\mu &\equiv& (\sinh\vartheta,0,0,\cosh\vartheta ) \, .
\label{eq:4vectors}
\ea
For a boost-invariant and cylindrically-symmetric system, one can simplify the basis vectors by identifying $\vartheta=\varsigma$ and $\varphi=\phi$ where $\varsigma$ and $\phi$ are the spacetime rapidity and the azimuthal angle, respectively.  In this case, the basis vectors (\ref{eq:basisvecs}) simplify to
\ba
u^\mu &=& (\cosh\theta_\perp \cosh\varsigma,\sinh\theta_\perp \cos\phi,\sinh\theta_\perp \sin\phi,\cosh\theta_\perp \sinh\varsigma) \, , \nonumber \\
X^\mu &=& (\sinh\theta_\perp \cosh\varsigma, \cosh\theta_\perp \cos\phi,\cosh\theta_\perp \sin\phi,\sinh\theta_\perp \sinh\varsigma) \, , \nonumber \\
Y^\mu &=&(0,-\sin\phi,\cos\phi,0)\, , \nonumber \\
Z^\mu &=&(\sinh\varsigma,0,0,\cosh\varsigma)\, .
\ea
In the case of a transversally-symmetric system, the transverse flow $u_\perp$ is absent, i.e. $\theta_\perp=0$, and, as a consequence, one has 
\ba
u^\mu &=& (\cosh\varsigma,0,0, \sinh\varsigma)\,,\nonumber \\
X^\mu &=& (0,\cos\phi,\sin\phi,0)\,,\nonumber \\
Y^\mu &=& (0,-\sin\phi,\cos\phi,0)\,,\nonumber \\
Z^\mu &=& (\sinh\varsigma,0,0,\cosh\varsigma).
\ea
Note that in the last case, $X^\mu$ and $Y^\mu$ are simply unit vectors pointing along the radial and azimuthal directions, respectively.
\subsection{Ellipsoidal form including bulk pressure degree of freedom}
\label{subsec:distribution-func}
In the non-conformal case, anisotropic hydrodynamics is defined through the introduction of an anisotropy tensor of the form \cite{Nopoush:2014pfa,Martinez:2012tu}
\be
\Xi^{\mu\nu} = u^\mu u^\nu + \xi^{\mu\nu} - \Delta^{\mu\nu} \Phi \, ,
\ee
where $u^\mu$ is four-velocity, $\xi^{\mu\nu}$ is a symmetric and traceless tensor, and $\Phi$ is associated with the bulk degree of freedom.  The quantities $u^\mu$, $\xi^{\mu\nu}$, and $\Phi$ are understood to be functions of spacetime and obey $u^\mu u_\mu = 1$, ${\xi^{\mu}}_\mu = 0$, ${\Delta^\mu}_\mu = 3$, and $u_\mu \xi^{\mu\nu} = 0$; therefore, one has ${\Xi^\mu}_\mu = 1 - 3 \Phi$.
At leading order in the anisotropic hydrodynamics expansion one assumes that the one-particle distribution function is of the form
\be
f(x,p) = f_{\rm iso}\!\left(\frac{1}{\lambda}\sqrt{p_\mu \Xi^{\mu\nu} p_\nu}\right) ,
\label{eq:genf}
\ee
where $\lambda$ has dimensions of energy and can be identified with the temperature only in the isotropic equilibrium limit ($\xi^{\mu\nu} = 0$ and $\Phi=0$).\footnote{Herein we assume that the chemical potential is zero.} 
We note that, in practice, $f_{\rm iso}$ need not be a thermal equilibrium distribution.  However, unless one expects there to be a non-thermal distribution at late times, it is appropriate to take $f_{\rm iso}$ to be a thermal equilibrium distribution function of the form $f_{\rm iso}(x) = f_{\rm eq}(x) = (e^x + a )^{-1}$,  where $a= \pm 1$ gives Fermi-Dirac or Bose-Einstein statistics, respectively, and $a=0$ gives Boltzmann statistics. From here on, we assume that the distribution is of Boltzmann form, i.e. $a=0$. 
 
\subsection{Dynamical Variables}
\label{subsec:dynamical-var}

Since the most important viscous corrections are to the diagonal components of the energy-momentum tensor, to good approximation one can assume that \mbox{$\xi^{\mu\nu} = {\rm diag}(0,{\boldsymbol \xi})$} with ${\boldsymbol \xi} \equiv (\xi_x,\xi_y,\xi_z)$ and $\xi^i_i=0$.  In this case, expanding the argument of the square root appearing on the right-hand side of Eq.~(\ref{eq:genf}) in the LRF gives
\be
f(x,p) =  f_{\rm eq}\!\left(\frac{1}{\lambda}\sqrt{\sum_i \frac{p_i^2}{\alpha_i^2} + m^2}\right) ,
\label{eq:fform}
\ee
where $i\in \{x,y,z\}$ and the scale parameters $\alpha_i$ are
\be
\alpha_i \equiv (1 + \xi_i + \Phi)^{-1/2} \, .
\label{eq:alphadef}
\ee  
Note that, for brevity, one can collect the three anisotropy parameters into vector $\boldsymbol\alpha \equiv (\alpha_x,\alpha_y,\alpha_z)$. In the isotropic equilibrium limit, where $\xi_i = \Phi = 0$ and $\alpha_i =1$, one has $p_\mu \Xi^{\mu\nu} p_\nu = (p \cdot u)^2 = E^2$  and $\lambda\rightarrow T$ and, therefore,
\be
f(x,p)=f_{\rm eq}\!\left(\frac{E}{T(x)}\right) .
\label{eq:feqform}
\ee
Out of the four anisotropy and bulk parameters there are only three independent ones. In practice, we use three variables $\alpha_i$ as the dynamical anisotropy parameters since, by using Eq.~(\ref{eq:alphadef}) and the tracelessness of $\xi^{\mu\nu}$, one can write $\Phi$ in terms of the anisotropy parameters, $\Phi = \frac{1}{3} \sum_i \alpha_i^{-2} - 1$.  In the transversally-homogeneous case, one has $\alpha_x=\alpha_y$ and, as a result, there are two independent anisotropy parameters.  Note that, for conformal systems, one has $\Phi=0$ and in this case there are then only two independent anisotropy parameters in 3+1d.

\subsection{Equation of state}
\label{subsec:eos}

Herein, we consider a system at finite temperature and zero chemical potential. At asymptotically high temperatures, the pressure of a gas of quarks and gluons approaches the Stefan-Boltzmann (SB) limit, ${\cal P}_{\rm SB}= {\cal E}_{\rm SB}/3 = N_{\rm dof} T^4 / \pi^2 = \pi^2T^4\left(N_c^2-1+\frac{7}{4}N_c N_f\right)/45$.  We will take $N_c=N_f=3$ in what follows.  At the temperatures probed in heavy-ion collisions there are important corrections to the SB limit and at low temperatures the relevant degrees of freedom change from quarks and gluons to hadrons.  The standard way to determine the QGP EoS is to use non-perturbative lattice calculations.  For this purpose, we use an analytic parameterization of lattice data for the QCD interaction measure (trace anomaly), $I_{\rm eq} = {\cal E}_{\rm eq} - 3 {\cal P}_{\rm eq}$, taken from the Wuppertal-Budapest collaboration \cite{Borsanyi:2010cj}
\be
\frac{I_{\rm eq}(T)}{T^4}=\bigg[\frac{h_0}{1+h_3 t^2}+\frac{f_0\big[\tanh(f_1t+f_2)+1\big]}{1+g_1t+g_2t^2}\bigg]\exp\!\Big(\!-\!\frac{h_1}{t}-\frac{h_2}{t^2}\Big) ,
\label{eq:I-func}
\ee
with $t\equiv T/(0.2 \; \rm GeV)$. For $N_f=2+1$ (2 light quarks and one heavy quark) the parameters are $h_0=0.1396$, $h_1=-0.1800$, $h_2=0.0350$, $f_0=2.76$, $f_1=6.79$, $f_2=-5.29$, $g_1=-0.47$, $g_2=1.04$, and $h_3=0.01$.

\begin{figure}[t]
\hspace{-6mm}
\includegraphics[width=.46\linewidth]{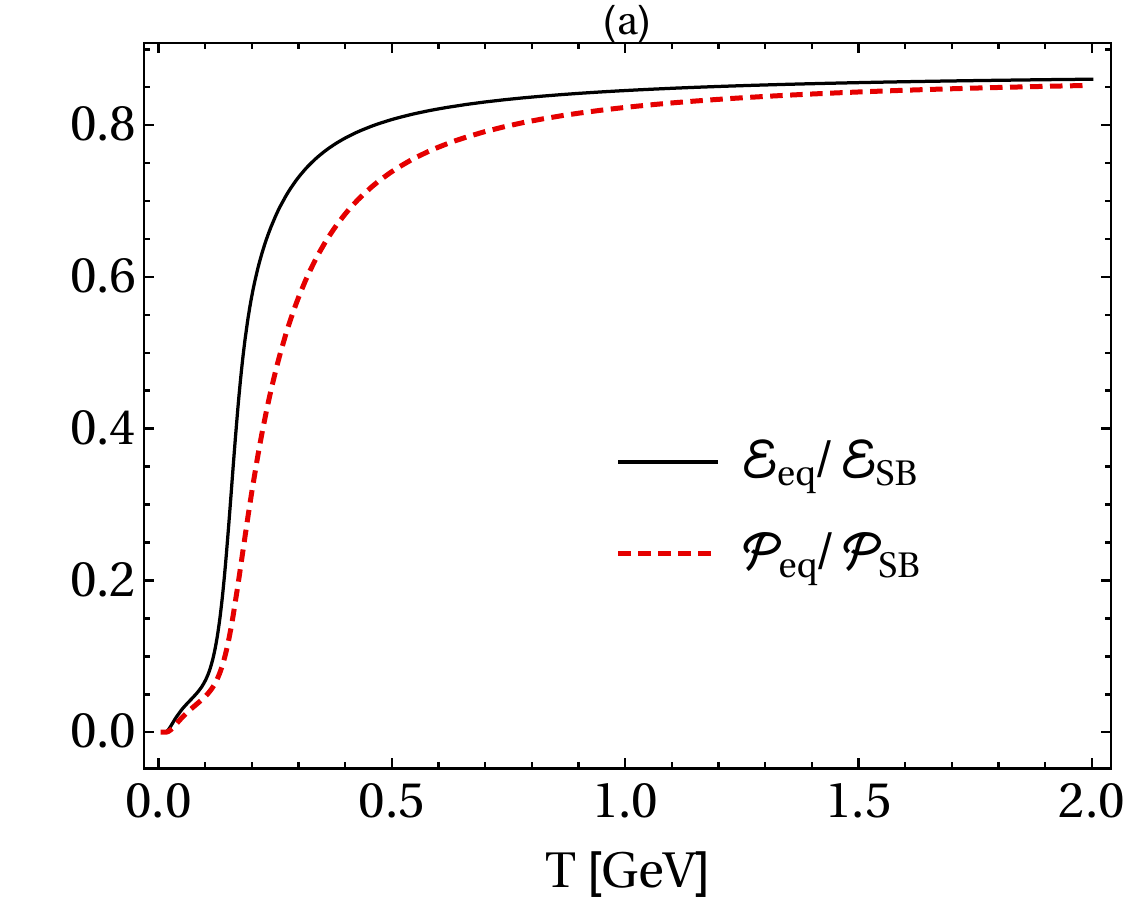}
\hspace{3mm}
\includegraphics[width=.48\linewidth]{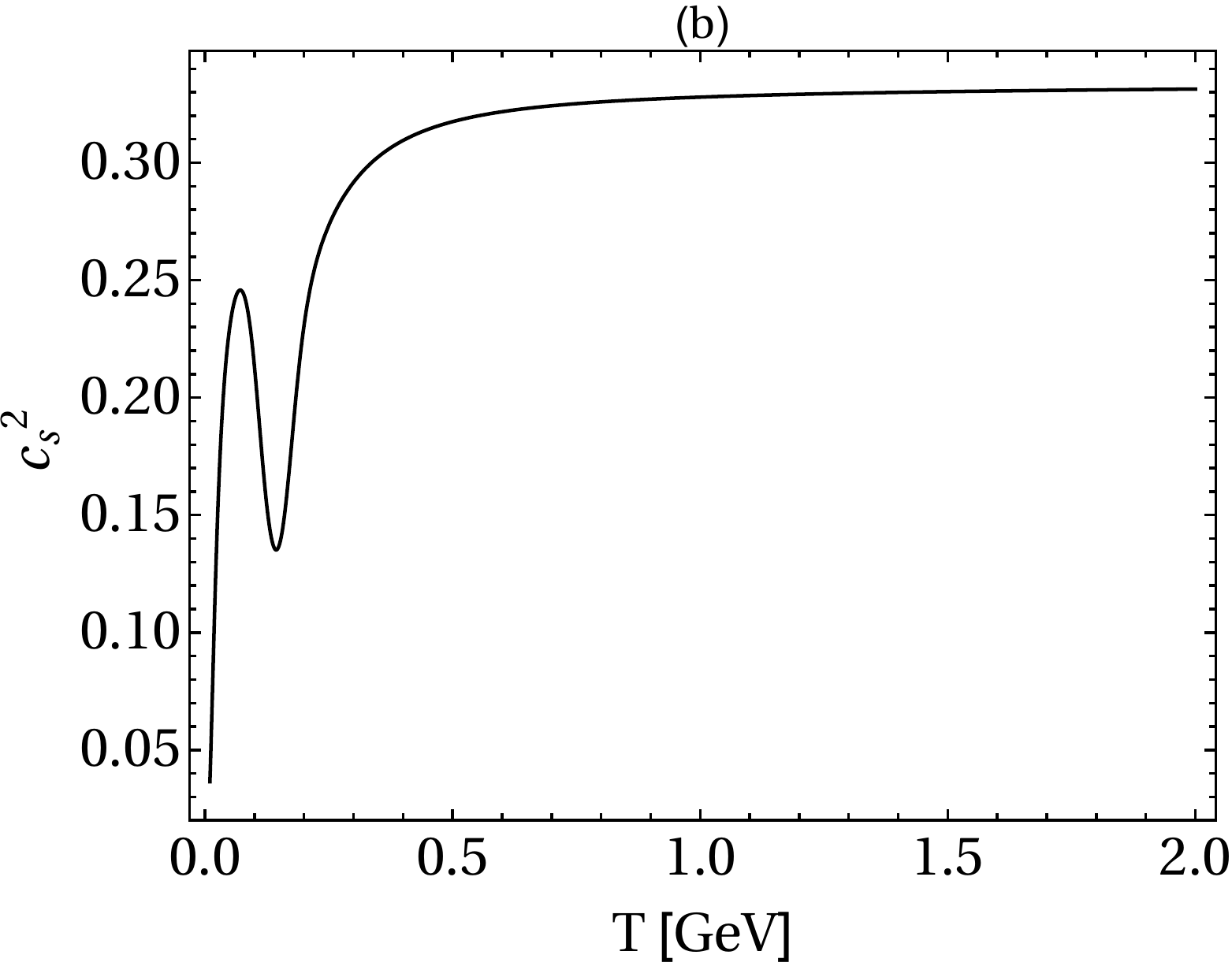}
\caption{Panel (a) shows the energy density and pressure scaled by their respective Stefan-Boltzmann limits as a function of temperature. Panel (b) shows the speed of sound squared as a function of temperature.}
\label{fig:eos}
\end{figure}

The pressure can be obtained from an integral of the interaction measure 
\be
\frac{{\cal P}_{\rm eq}(T)}{T^4}=\int_0^T \frac{dT}{T}\frac{I_{\rm eq}(T)}{T^4} \, ,
\label{eq:P-func}
\ee
where we have assumed ${\cal P}_{\rm eq}(T=0)=0$.  Having $P_{\rm eq}(T)$, one can obtain the energy density ${\cal E}_{\rm eq}$ using ${\cal E}_{\rm eq}(T) = 3 {\cal P}_{\rm eq}(T) + I_{\rm eq}(T)$.  In the limit $T\rightarrow\infty$, the system tends to the ideal limit as expected.\footnote{In the original parametrization presented in Ref.~\cite{Borsanyi:2010cj} the authors used $h_3=0$, however, as pointed out in Ref.~\cite{Nopoush:2015yga}, choosing $h_3=0$ gives the wrong high temperature limit.} The temperature dependence of the resulting equilibrium energy density, pressure, and speed of sound squared ($c_s^2 = \partial {\cal P}_{\rm eq}/\partial{\cal E}_{\rm eq}$) are shown in the two panels of Fig.~\ref{fig:eos}.  

\subsubsection*{Method 1:  Standard equation of state}

In the standard approach for imposing a realistic EoS in anisotropic hydrodynamics, one derives the necessary equations in the conformal limit and exploits the conformal multiplicative factorization of the components of the energy-momentum tensor~\cite{Martinez:2010sc,Florkowski:2010cf}.  With this method, one relies on the assumption of factorization even in the non-conformal (massive) case.  Such an approach is justified by the smallness of the corrections to factorization in the massive case in the near-equilibrium limit~\cite{Nopoush:2015yga}.  For details concerning this method, we refer the reader to Refs.~\cite{Ryblewski:2012rr,Nopoush:2015yga}.  Although this method is relatively straightforward to implement, it is only approximate since for non-conformal systems there is no longer exact multiplicative factorization of the components of the energy-momentum tensor.  This introduces a theoretical uncertainty which is difficult to quantitatively estimate.

\subsubsection*{Method 2:  Quasiparticle equation of state}

Since the standard method is only approximate, one would like to find an alternative method for imposing a realistic equation of state in an anisotropic system that can be applied for non-conformal systems.  In order to accomplish this goal, we  implement the realistic EoS detailed above by assuming that the QGP can be described as an ensemble of massive quasiparticles with temperature-dependent masses.  As is well-known from the literature \cite{gorenstein1995gluon}, one cannot simply substitute temperature-dependent masses into the thermodynamic functions obtained with constant masses because this would violate thermodynamic consistency.  For an equilibrium system, one can ensure thermodynamic consistency by adding a background contribution to the energy-momentum tensor, i.e.
\be
T^{\mu\nu}_{\rm eq} = T^{\mu\nu}_{\rm kinetic,eq} + g^{\mu\nu} B_{\rm eq}  \, ,
\ee
with $B_{\rm eq}\equiv B_{\rm eq}(T)$ being the additional background contribution.  The kinetic contribution to the energy momentum tensor is given by
\be
T^{\mu\nu}_{\rm kinetic,eq} = \intdP \, p^\mu p^\nu f_{\rm eq}(x,p) \, .
\ee

For an equilibrium Boltzmann gas, the number and entropy densities are unchanged, while, due to the additional background contribution, the energy density and pressure are shifted by $+B_{\rm eq}$ and $-B_{\rm eq}$, respectively, giving
\ba
n_{\rm eq}(T,m) &=& 4 \pi \tilde{N} T^3 \, \hat{m}_{\rm eq}^2 K_2\left( \hat{m}_{\rm eq}\right) , \label{eq:neq} \\
{\cal S}_{\rm eq}(T,m) &=&4 \pi \tilde{N} T^3 \, \hat{m}_{\rm eq}^2 \Big[4K_2\left( \hat{m}_{\rm eq}\right)+\hat{m}_{\rm eq}K_1\left( \hat{m}_{\rm eq}\right)\Big] ,
\label{eq:Seq} \\
{\cal E}_{\rm eq}(T,m) &=& 4 \pi \tilde{N} T^4 \, \hat{m}_{\rm eq}^2
 \Big[ 3 K_{2}\left( \hat{m}_{\rm eq} \right) + \hat{m}_{\rm eq} K_{1} \left( \hat{m}_{\rm eq} \right) \Big]+B_{\rm eq} \, , 
\label{eq:Eeq} \\
 {\cal P}_{\rm eq}(T,m) &=& 4 \pi \tilde{N} T^4 \, \hat{m}_{\rm eq}^2 K_2\left( \hat{m}_{\rm eq}\right)-B_{\rm eq} \, ,
\label{eq:Peq}
\ea
where $\hat{m}_{\rm eq} = m/T$ with $m$ implicitly depending on the temperature from here on.  In order to fix $B_{\rm eq}$, one can require, for example, the thermodynamic identity
\be
T {\cal S}_{\rm eq} = {\cal E}_{\rm eq} + {\cal P}_{\rm eq} = T \frac{\partial P_{\rm eq}}{\partial T} \, ,
\label{eq:thermoid}
\ee
be satisfied.  Using Eqs.~(\ref{eq:Eeq}), (\ref{eq:Peq}), and (\ref{eq:thermoid}) one obtains 
\ba
\frac{dB_{\rm eq}}{dT} &=& - \frac{1}{2} \frac{dm^2}{dT} \intdP \, f_{\rm eq}(x,p) \nonumber \\  
&=& -4\pi \tilde{N}m^2 T K_1(\hat{m}_{\rm eq}) \frac{dm}{dT} \, .
\label{eq:BM-matching-eq-1}
\ea
If the temperature dependence of $m$ is known, then Eq.~(\ref{eq:BM-matching-eq-1}) can be used to determine $B_{\rm eq}$.

\begin{figure}[t]
\hspace{-6mm}
\includegraphics[width=0.46\linewidth]{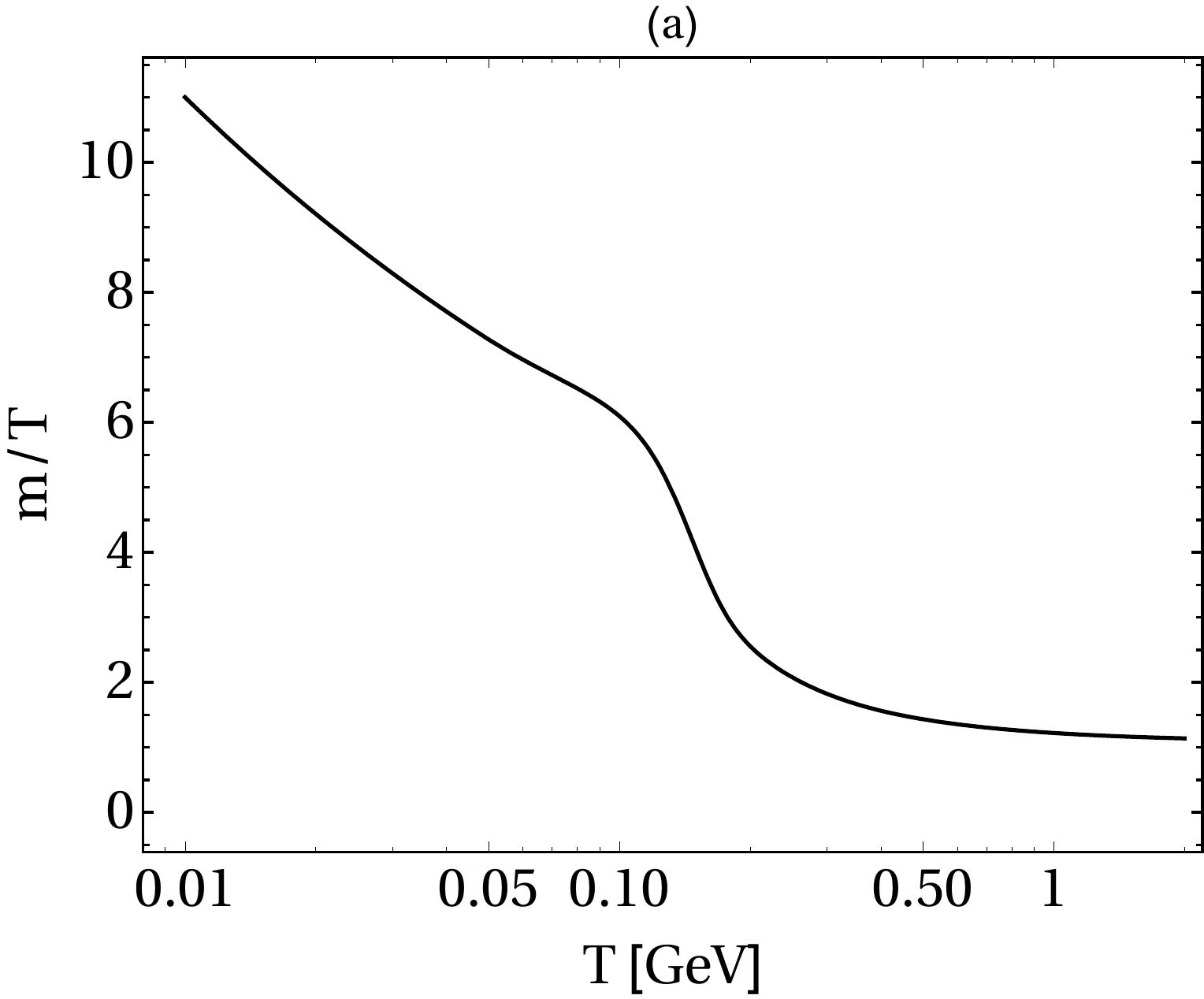}
$\;\;$
\includegraphics[width=0.48\linewidth]{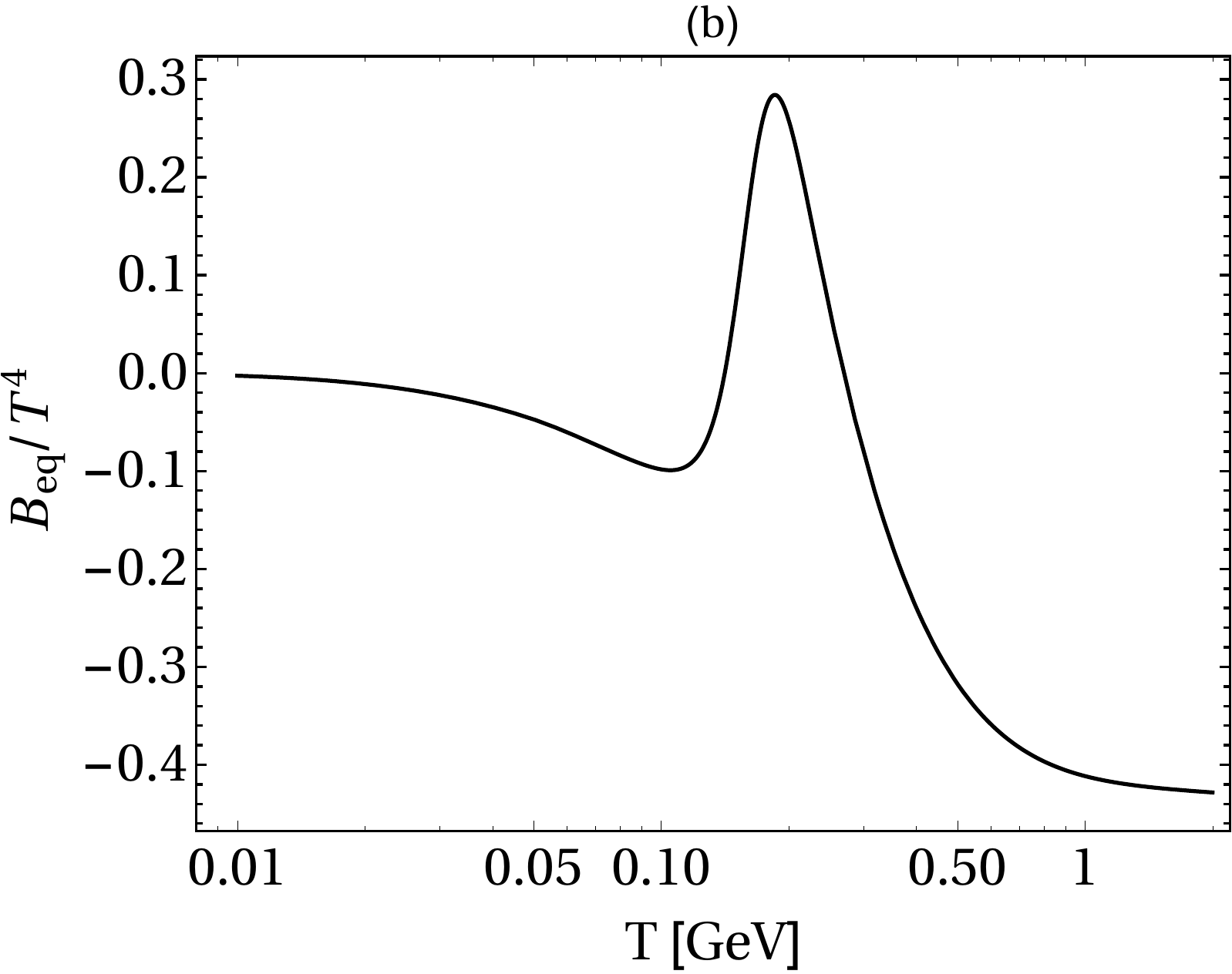}
\caption{In panel (a) we plot the temperature dependence of the quasiparticle mass scaled by the temperature obtained using Eq.~(\ref{eq:meq}).  In panel (b) we plot the temperature dependence of the background term $B_{\rm eq}$ scaled by the temperature obtained using (\ref{eq:BM-matching-eq-1}).}
\label{fig:eos2}
\end{figure}

In order to determine $m$, one can use the thermodynamic identity
\be
{\cal E}_{\rm eq}+{\cal P}_{\rm eq}=T{\cal S}_{\rm eq} = 4 \pi \tilde{N} T^4 \, \hat{m}_{\rm eq}^3 K_3\left( \hat{m}_{\rm eq}\right) .
\label{eq:meq}
\ee
Using the lattice parameterization (\ref{eq:I-func}) to compute the equilibrium energy density and pressure, one can numerically solve for $m(T)$.  In Fig.~\ref{fig:eos2}a, we plot the resulting solution for $m/T$ as a function of the temperature.  Once $m$ is determined using Eq.~(\ref{eq:meq}), one can solve Eq.~(\ref{eq:BM-matching-eq-1}) subject to the boundary condition $B_{\rm eq}(T=0) = 0$ to find $B_{\rm eq}(T)$. We note that, using this method, one can exactly reproduce the lattice results for energy density, pressure, and entropy density.  In Fig.~\ref{fig:eos2}b, we plot the resulting solution for the normalized quantity $B_{\rm eq}(T)/T^4$  as a function of the temperature.  

\section{Boltzmann equation and its moments}
\label{sec:boltzman}

In this paper, we derive the necessary hydrodynamical equations by taking the moments of Boltzmann equation.  In what follows, we specialize to the case that the collisional kernel is given by the relaxation-time approximation (RTA), however, the general methods presented here can be applied to any collisional kernel.  If the particles that comprise the system have temperature-independent masses then the Boltzmann equation is of the form
\be
p^\mu \partial_\mu f = -{\cal C}[f]\, .
\label{eq:boltzmanneq}
\ee
The function ${\cal C}[f]$ at right-hand side of the equation is the collisional kernel containing all interactions involved in the dynamics. In RTA, one has
\be
{\cal C}[f]=\frac{p^\mu u_\mu}{\tau_{\rm eq}}(f-f_{\rm eq})\,.
\label{eq:RTA}
\ee
In this relation, $f_{\rm eq}$ denotes the equilibrium one-particle distribution function (\ref{eq:feqform}) and $\tau_{\rm eq}$ is the relaxation time which can depend on spacetime but which we assume to be momentum-independent.  To obtain a realistic model for $\tau_{\rm eq}$, which is valid for massive systems, one can relate $\tau_{\rm eq}$ to the shear viscosity to entropy density ratio.  For a massive system, one has \cite{anderson1974relativistic,Czyz:1986mr}
\be
\eta(T)=\frac{\tau_{\rm eq}(T) {\cal P}_{\rm eq}(T)}{15}\kappa(\hat{m}_{\rm eq})\,.
\ee
In this formula the function $\kappa(x)$ is defined as
\be 
\kappa(x)\equiv x^3 \bigg[\frac{3}{x^2}\frac{K_3(x)}{K_2(x)}-\frac{1}{x}+\frac{K_1(x)}{K_2(x)}-\frac{\pi}{2}\frac{1-xK_0(x)L_{-1}(x)-xK_1(x)L_0(x)}{K_2(x)}\bigg] ,
\ee
where $K_n(x)$ are modified Bessel functions of second kind and $L_n(x)$ are modified Struve functions. Assuming that the ratio of the shear viscosity to entropy density, $\eta/{\cal S}_{\rm eq}\equiv \bar{\eta}$, is held fixed during the evolution and using the thermodynamic relation ${\cal E}_{\rm eq}+{\cal P}_{\rm eq}=T{\cal S}_{\rm eq}$ one obtains
\be 
\tau_{\rm eq}(T)=\frac{15 \bar{\eta}}{\kappa(\hat{m}_{\rm eq})T}\bigg(1+\frac{{\cal E}_{\rm eq}(T)}{{\cal P}_{\rm eq}(T)}\bigg) .
\label{eq:teq}
\ee
Note that, in the massless limit, $m\rightarrow 0$, one has $\kappa(\hat{m}_{\rm eq})\rightarrow12$, giving 
\be
\tau_{\rm eq}(T)= \frac{5 \eta}{4{\cal P}_{\rm eq}(T)} \, .
\label{eq:teq0}
\ee

\subsection{Effective Boltzmann Equation}
\label{subsec:thermal-boltzmann}
If the quasiparticles have a temperature-dependent mass, one has to generalize the Boltzmann equation in order to take into account gradients in the mass. Generally, the Boltzmann equation for on-shell quasiparticles can be written as~\cite{Jeon:1995zm}
\be
\left(\frac{\partial}{\partial t} + \frac{\partial E}{\partial \bf{p}} \cdot
\frac{\partial}{\partial \bf{x}} - \frac{\partial E}{\partial \bf{x}} \cdot
\frac{\partial}{\partial \bf{p}}\right) f(x,p) = \left(\frac{\partial f}{\partial t}\right)_{\rm coll}
 \,,
\label{eq:boltz1}
\ee
where the ``external force term'' $-dE/d\bf{x}$ does not vanish in the thermal mass case since the particle energy depends on the mass and hence on the local temperature of the system.   As a result, the temperature-dependence of the mass acts as an external force in the dynamics.  Using the on-shell energy relation $E\equiv\sqrt{{\bf p}^2+m^2}$ and defining the collisional kernel as
\be
\mathcal{C}[f]\equiv-E\left(\frac{\partial f}{\partial t}\right)_{\rm coll} ,
\ee
in covariant form one has 
\be
p^\mu \partial_\mu f+\frac{1}{2}\partial_i m^2\partial^i_{(p)} f=-\mathcal{C}[f]\,,
\label{eq:boltz2}
\ee
where $p^\mu\equiv(\sqrt{{\bf p}^2+m^2},\bf p)$ is the on-shell momentum four-vector, $i$ is a spatial coordinate index, and
$\partial^i_{(p)}\equiv -\partial/\partial p^i$.  Note that the extra term, $(\partial_i m^2/2)\partial^i_{(p)} f$, corresponds precisely to the result obtained from derivation of the Boltzmann equation using quantum field theoretical methods~\cite{Berges:2005md}.  

We mention that an alternative way of deriving the effective Boltzmann equation above can be found in a recent paper of Romatschke \cite{Romatschke:2011qp}. In this paper, a general Boltzmann equation for off-shell particles (with constant mass) is first derived using the evolution of a single particle distribution function along eight-dimensional phase space geodesics, where the possibility of curved spaces is taken into account using geometrical covariant derivatives.  Then, by adding a temperature-dependent background term to $T^{\mu\nu}$, temperature-dependent masses are taken into account in a way that guarantees both thermodynamic consistency in the equilibrium limit and energy-momentum conservation, in general. Finally, the on-shell version of the effective Boltzmann equation for quasiparticles with a temperature-dependent mass in a general curved space time is obtained. The flat spacetime limit of the effective Boltzmann equation derived by Romatschke is the same as Eq.~(\ref{eq:boltz2}).

\subsection{Moments of Boltzmann Equation}
\label{subsec:boltzmann-moments}
If one is interested in the evolution of the bulk properties of a system, one can use low-order moments of the Boltzmann equation.  By calculating moments of Boltzmann equation one obtains evolution equations for tensors of different ranks, with the first moment giving an evolution equation for the energy-momentum tensor and the second-moment describing the evolution of a rank three tensor.
Taking the zeroth, first, and second moments of Boltzmann equation gives, respectively 
\ba
\partial_\mu J^\mu&=&-\intdP \, {\cal C}[f]\, , \label{eq:J-conservation} \\
\partial_\mu T^{\mu\nu}&=&-\intdP \, p^\nu {\cal C}[f]\, , \label{eq:T-conservation} \\
\partial_\mu {\cal I}^{\mu\nu\lambda}- J^{(\nu} \partial^{\lambda)} m^2 &=&-\intdP \, p^\nu p^\lambda{\cal C}[f]\, \label{eq:I-conservation},   
\ea 
where the particle four-current $J^\mu$, energy-momentum tensor $T^{\mu\nu}$, and the rank-three tensor ${\cal I}^{\mu\nu\lambda}$ are given by
\ba
J^\mu &\equiv& \intdP \, p^\mu f(x,p)\, , \label{eq:J-int} \\
T^{\mu\nu}&\equiv& \intdP \, p^\mu p^\nu f(x,p)+B g^{\mu\nu}, \label{eq:T-int}\\
{\cal I}^{\mu\nu\lambda} &\equiv& \intdP \, p^\mu p^\nu p^\lambda  f(x,p) \, .
\label{eq:I-int}
\ea
We note that we have introduced the non-equilibrium background field $B\equiv B({\boldsymbol\alpha},\lambda)$, which is the analogue of the equilibrium background $B_{\rm eq}$ in order to guarantee that the correct equilibrium limit of $T^{\mu\nu}$ is obtained.  In the process of the derivation one finds that, in order to write the energy momentum conservation in the form given in Eq.~(\ref{eq:T-conservation}), there must be a differential equation relating $B$ and the thermal mass
\be
\partial_\mu B = -\frac{1}{2} \partial_\mu m^2 \intdP  f(x,p)\,.
\label{eq:BM-matching}
\ee
In practice, one can use (\ref{eq:BM-matching}) to write the derivative of $B$ with respect to any variable in terms of the derivative of the thermal mass times the $E^{-1}$ moment of the non-equilibrium distribution function.

\section{Bulk variables}
\label{sec:bulk-var}

In this section, bulk variables, i.e. number density, energy density, and the pressures, are calculated by taking the projections of $J^\mu$ and $T^{\mu\nu}$.

\subsection{Particle current 4-vector}
\label{subsec:4current}
The particle current four-vector $J^\mu\equiv(n,\bf{J})$ is defined in Eq.~(\ref{eq:J-int}). One can expand $J^\mu$ using the basis vectors as
\be
J^\mu=n u^\mu+J_x X^\mu+J_y Y^\mu+J_z Z^\mu\, .
\label{eq:J-exp}
\ee
Using Eqs.~(\ref{eq:fform}) and (\ref{eq:J-int}) one has
\be
J^\mu=(n,{\bf 0})=nu^\mu \, ,
\label{eq:J-exp2}
\ee
where $n=\alpha n_{\rm eq}(\lambda,m)$ and $\alpha\equiv\alpha_x \alpha_y \alpha_z$.

\subsection{Energy-Momentum Tensor}  
\label{subsec:T-tensor}

The energy-momentum tensor $T^{\mu\nu}$ is defined in Eq.~(\ref{eq:T-int}). Expanding it using the basis vectors one obtains
\be
T^{\mu\nu}={\cal E}u^\mu u^\nu+{\cal P}_x X^\mu X^\nu+{\cal P}_y Y^\mu Y^\nu+{\cal P}_z Z^\mu Z^\nu \, .
\label{eq:T-expan}
\ee
Using Eqs.~(\ref{eq:fform}), (\ref{eq:T-int}), and (\ref{eq:T-expan}) and taking projections of $T^{\mu\nu}$ one can obtain the energy density and the components of pressure
\ba
{\cal E} &=& {\cal H}_3({\boldsymbol\alpha},\hat{m}) \, \lambda^4+B \, ,\nonumber \\
{\cal P}_x &=& {\cal H}_{3x}({\boldsymbol\alpha},\hat{m}) \, \lambda^4-B \, ,\nonumber \\
{\cal P}_y &=& {\cal H}_{3y}({\boldsymbol\alpha},\hat{m}) \, \lambda^4-B \, ,\nonumber \\
{\cal P}_z &=& {\cal H}_{3L}({\boldsymbol\alpha},\hat{m}) \, \lambda^4-B \, ,
\ea
where $\hat{m} \equiv m/\lambda$.  In the transversally-symmetric case one has ${\cal P}_T\equiv {\cal P}_x={\cal P}_y$ and ${\cal P}_L\equiv{\cal P}_z$ and Eq.~(\ref{eq:T-expan}) simplifies to
\be
T^{\mu \nu} = \left( {\cal E}+ {\cal P}_T \right)  u^\mu u^\nu
- {\cal P}_T g^{\mu\nu}+\left( {\cal P}_L - {\cal P}_T \right) Z^\mu Z^\nu \, ,
\label{eq:T-expan-trans}
\ee
where 
\ba
{\cal E} &=& \tilde{{\cal H}}_3({\boldsymbol\alpha},\hat{m}) \, \lambda^4+B \, ,\nonumber \\
{\cal P}_T &=& \tilde{{\cal H}}_{3T}({\boldsymbol\alpha},\hat{m}) \, \lambda^4-B \, ,\nonumber \\
{\cal P}_L &=& \tilde{{\cal H}}_{3L}({\boldsymbol\alpha},\hat{m}) \, \lambda^4-B \,.
\label{eq:E-P-trans}
\ea
The various ${\cal H}$-functions appearing above are defined in App.~\ref{subapp:h-functions-1}.

\section{Dynamical equations}
\label{sec:dynamical-eqs}

In order to obtain the dynamical equations from Eqs.~(\ref{eq:J-conservation})-(\ref{eq:I-conservation}), one needs the tensor decomposition of $J^\mu$, $T^{\mu\nu}$, and ${\cal I}^{\mu\nu\lambda}$ using the basis vectors. Herein the general 3+1d equations for a system with temperature-dependent masses are obtained in the RTA. We then simplify to the case of 0+1d transversally-symmetric case by the taking necessary limits. In what follows, the convective derivatives $D_\alpha$ and divergences $\theta_\alpha$, with $\alpha\in\{u,x,y,z\}$, are defined in App.~\ref{app:identities}.

\subsection{Zeroth moment}
\label{subsec:0th-moment}

The evolution equation for the particle four-current (\ref{eq:J-conservation}) in the RTA is
\be
\partial_\mu J^\mu=\frac{1}{\tau_{\rm eq}}(n_{\rm eq}-n) \, .
\label{eq:0th-mom-gen}
\ee
Using Eq.~(\ref{eq:J-exp2}) one has
\be
D_un+n\theta_u=\frac{1}{\tau_{\rm eq}}(n_{\rm eq}-n) \, .
\label{eq:0th-mom}
\ee
In the case of 0+1d, this simplifies to
\be
\partial_\tau n+\frac{n}{\tau}=\frac{1}{\tau_{\rm eq}}(n_{\rm eq}-n) \, .
\label{eq:0th-mom-trans}
\ee
%
\subsection{First Moment}
\label{subsec:1st-moment}

The conservation of energy and momentum is enforced by $\partial_\mu T^{\mu\nu}=0$.  This requires that both the left and right hand sides of Eq.~(\ref{eq:T-conservation}) vanish.  The vanishing of the right-hand side of this equation results in a constraint equation that can be used to write $T$ in terms of the non-equilibrium microscopic parameters ${\boldsymbol\alpha}$ and $\lambda$.  Using (\ref{eq:fform}), (\ref{eq:feqform}), and (\ref{eq:RTA}) one obtains ${\cal E}_{\rm kinetic} = {\cal E}_{\rm kinetic,eq}$, or more explicitly
\be
\tilde{\cal H}_3 \lambda^4 = \tilde{\cal H}_{3,\rm eq} T^4.
 \label{eq:matching}
\ee

Turning to the left hand side, using Eq.~(\ref{eq:T-expan}) and taking $U$-, $X$-, $Y$-, and $Z$-projections, one obtains four independent equations
\ba
D_u{\cal E}+{\cal E}\theta_u+ {\cal P}_x u_\mu D_xX^\mu+ {\cal P}_y u_\mu D_yY^\mu +{\cal P}_z u_\mu D_zZ^\mu &=&0\, , \nonumber\\
D_x {\cal P}_x+{\cal P}_x\theta_x -{\cal E}X_\mu D_uu^\mu -{\cal P}_y X_\mu D_yY^\mu - {\cal P}_z X_\mu D_zZ^\mu &=& 0\,, \nonumber\\
D_y {\cal P}_y+{\cal P}_y \theta_y-{\cal E}Y_\mu D_uu^\mu -{\cal P}_x Y_\mu D_xX^\mu - {\cal P}_z Y_\mu D_zZ^\mu  &=& 0\,, \nonumber\\
D_z {\cal P}_z+{\cal P}_z \theta_z-{\cal E}Z_\mu D_uu^\mu- {\cal P}_x Z_\mu D_xX^\mu - {\cal P}_y Z_\mu D_yY^\mu &=& 0\,.
\label{eq:1st-mom-gen}
\ea
In the 0+1d case, using ${\cal P}_T\equiv {\cal P}_x={\cal P}_y$ and ${\cal P}_L\equiv{\cal P}_z$ and taking the appropriate limits, as explained in App.~\ref{app:identities}, one can simplify Eqs.~(\ref{eq:1st-mom-gen}) to
\ba
\partial_\tau {\cal E}&=&-\frac{{\cal E+P}_L}{\tau}\,,\label{eq:1st-mom-u}\\
\partial_r {\cal P}_T&=& \partial_\phi {\cal P}_T =  \partial_\varsigma {\cal P}_L = 0 \,. \label{eq:1st-mom-zero}
\ea
Eqs.~(\ref{eq:1st-mom-zero}) are consequences of boost invariance and transverse homogeneity in the 0+1d case and, as a result, the only independent dynamical equation is Eq.~(\ref{eq:1st-mom-u}).

\subsection{Second moment}
\label{subsec:2nd-moment}
The second moment of Boltzmann equation in the RTA is
\be
\partial_\mu {\cal I}^{\mu\nu\lambda}-J^{(\nu} \partial^{\lambda)} m^2= \frac{1}{\tau_{\rm eq}}(u_\mu {\cal I}^{\mu\nu\lambda}_{\rm eq}-u_\mu {\cal I}^{\mu\nu\lambda})\,,
\label{eq:2moment}
\ee
where ${\cal I}^{\mu\nu\lambda}_{\rm eq}$ can be obtained from Eq.~(\ref{eq:I-int}) by taking $f \rightarrow f_{\rm eq}$.  For a distribution function of the form specified in Eq.~(\ref{eq:fform}), the only non-vanishing terms in Eq.~(\ref{eq:I-int}) are those with an even number of similar spatial index. As a result, one can expand ${\cal I}^{\mu\nu\lambda}$ over the basis vectors as
\ba
{\cal I} &=&\, {\cal I}_u \left[ u\otimes u \otimes u\right] 
\nonumber \\
&+&\, {\cal I}_x \left[ u\otimes X \otimes X +X\otimes u \otimes X + X\otimes X \otimes u\right] 
\nonumber \\ 
&+&\,  {\cal I}_y  \left[ u\otimes Y \otimes Y +Y\otimes u \otimes Y + Y\otimes Y \otimes u\right]
\nonumber \\
&+&\, {\cal I}_z \left[ u\otimes Z \otimes Z +Z\otimes u \otimes Z + Z\otimes Z \otimes u\right] .
 \label{eq:Theta}
\ea
Evaluating the necessary integrals using the distribution function (\ref{eq:fform}), one finds
\ba
{\cal I}_u &=& \Big(\sum_i \alpha_i^2\Big) \alpha \, {\cal I}_{\rm eq}(\lambda,m) + \alpha m^2 n_{\rm eq}(\lambda,m) \, ,
\\
{\cal I}_i &=& \alpha \, \alpha_i^2 \, {\cal I}_{\rm eq}(\lambda,m) \, ,
\label{eq:I-i}
\ea
where
\be
{\cal I}_{\rm eq}(\lambda,m) =  4 \pi {\tilde N} \lambda^5 \hat{m}^3 K_3(\hat{m}) \, .
\label{eq:Ieq}
\ee
Note that, in general, one has ${\cal I}_u - \sum_i {\cal I}_i = \alpha m^2 n_{\rm eq}(\lambda,m)$ and $\lim_{m \rightarrow 0} {\cal I}_u = \sum_i {\cal I}_i$. 
Expanding Eq.~(\ref{eq:2moment}) and taking its $uu$-, $XX$-, $YY$-, and $ZZ$-projections gives
\ba
D_u {\cal I}_u + {\cal I}_u \theta_u + 2 {\cal I}_x u_\mu D_x X^\mu+ 2 {\cal I}_y u_\mu D_y Y^\mu+ 2 {\cal I}_z u_\mu D_z Z^\mu
-nD_u m^2
&=& \frac{1}{\tau_{\rm eq}} ( {\cal I}_{u,\rm eq} - {\cal I}_u ) \, , \;\;\;\; \label{eq:uu} \\
D_u {\cal I}_x + {\cal I}_x (\theta_u + 2 u_\mu D_x X^\mu)
&=& \frac{1}{\tau_{\rm eq}} ( {\cal I}_{\rm eq} - {\cal I}_x ) \, , \label{eq:xx} \\
D_u {\cal I}_y + {\cal I}_y (\theta_u + 2 u_\mu D_y Y^\mu)
&=& \frac{1}{\tau_{\rm eq}} ( {\cal I}_{\rm eq} - {\cal I}_y ) \, , \label{eq:yy}\\
D_u {\cal I}_z + {\cal I}_z (\theta_u + 2 u_\mu D_z Z^\mu)
&=& \frac{1}{\tau_{\rm eq}} ( {\cal I}_{\rm eq} - {\cal I}_z ) \, . \label{eq:zz} 
\ea
Also, taking $uX$-, $uY$-, and $uZ$-projections one can find
\ba
D_x {\cal I}_x+{\cal I}_x \theta_x+({\cal I}_x+{\cal I}_u) u_\mu D_u X^\mu -{\cal I}_y X_\mu D_y Y^\mu-{\cal I}_z X_\mu D_z Z^\mu-\frac{1}{2}nD_xm^2=0\,, \label{eq:ux} \\
D_y{\cal I}_y+{\cal I}_y \theta_y+({\cal I}_y+{\cal I}_u) u_\mu D_uY^\mu -{\cal I}_x Y_\mu D_x X^\mu-{\cal I}_z Y_\mu D_z Z^\mu-\frac{1}{2}nD_ym^2=0\,, \label{eq:uy}\\
D_z{\cal I}_z+{\cal I}_z \theta_z+({\cal I}_z+{\cal I}_u) u_\mu D_u Z^\mu -{\cal I}_x Z_\mu D_x X^\mu-{\cal I}_y Z_\mu D_y Y^\mu-\frac{1}{2}nD_zm^2=0\, , \label{eq:uz}
\ea
and finally projecting with $XY$, $XZ$, and $YZ$ gives
\ba
{\cal I}_x(Y_\mu D_u X^\mu+Y_\mu D_x u^\mu)+ {\cal I}_y (X_\mu D_u Y^\mu+X_\mu D_y u^\mu)&=&0\,, \label{eq:xy}\\
{\cal I}_x (Z_\mu D_u X^\mu+Z_\mu D_x u^\mu)+ {\cal I}_z (X_\mu D_u Z^\mu+X_\mu D_z u^\mu)&=&0\,, \label{eq:xz}\\
{\cal I}_y (Z_\mu D_u Y^\mu+Z_\mu D_y u^\mu)+ {\cal I}_z (Y_\mu D_u Z^\mu+Y_\mu D_z u^\mu)&=&0\,.  \label{eq:yz}
\ea

It can be shown that Eq.~(\ref{eq:uu}) is not independent. One can subtract the sum of Eqs.~(\ref{eq:xx})-(\ref{eq:zz}) from it to obtain
\be
m^2(D_un+n\theta_u)=\frac{m^2}{\tau_{\rm eq}}(n_{\rm eq}-n)\,.
\ee
This equation is the same as Eq.~(\ref{eq:0th-mom}) for non-vanishing mass.
In the 0+1d case, one has ${\cal I}_x={\cal I}_y$ and Eqs.~(\ref{eq:xx})-(\ref{eq:zz}) simplify to
\ba
\partial_\tau \log{\cal I}_x+\frac{1}{\tau} &=&\frac{1}{\tau_{\rm eq}}\Big(\frac{{\cal I}_{\rm eq}}{{\cal I}_x}-1\Big) ,\label{eq:xx-trans}\\
\partial_\tau  \log{\cal I}_z+\frac{3}{\tau} &=&\frac{1}{\tau_{\rm eq}}\Big(\frac{{\cal I}_{\rm eq}}{{\cal I}_z}-1\Big) . \label{eq:zz-trans}
\ea
Finally, we note that Eqs.~(\ref{eq:ux})-(\ref{eq:yz}) are trivially satisfied in the 0+1d case.

\subsection{Selection of relevant equations of motion}

For the 0+1d case, we need four equations for the four independent parameters, $\lambda, T, \alpha_x,\alpha_z$. Using the equations derived thus far up to the second moment of Boltzmann equation, we have five independent equations.  Herein, we use the equations obtained solely from the first and second moments of the Boltzmann equation which give Eqs.~(\ref{eq:matching}), (\ref{eq:1st-mom-u}), (\ref{eq:xx-trans}), and (\ref{eq:zz-trans}).\footnote{For the ``quasiparticle EoS'' case one obtains quite similar results if one instead uses the equation obtained from the zeroth-moment (\ref{eq:0th-mom-trans}); however, in the ``standard EoS'' case, one finds that using the zeroth moment equation (\ref{eq:0th-mom-trans}) results in solutions that do not approach the isotropic equilibrium limit at late times.}

\section{0+1d dynamical equations}
\label{sec:0p1}

In this section, we present the dynamical equations for the ``quasiparticle EoS'' and the ``standard EoS'' cases.  For simplicity, we present only the 0+1d case herein.  We postpone the 3+1d numerical comparisons to a future work.

\subsection{Quasiparticle equation of state}
\label{sec:thermal-mass}

One potential complication encountered when using temperature-dependent masses is that the first moment equation will involve the background contribution $B$ and its proper-time derivative, since on the left-hand side of (\ref{eq:1st-mom-u}) one has the total energy density which includes the background contribution.  In practice, however, all derivatives of $B$ can be written in terms of derivatives of $m$ using Eq.~(\ref{eq:BM-matching}).  For the 0+1d case, we only need the proper-time derivative of $B$.  Taking the distribution function to be of the form (\ref{eq:fform}) and using Eq.~(\ref{eq:BM-matching}) one obtains
\be
\partial_\tau B = - \frac{\lambda^2 }{2} \tilde{\cal H}_{3B}({\boldsymbol\alpha},\hat{m}) \, \partial_\tau m^2  \, .
\label{eq:B}
\ee
In this way, all proper-time derivatives of $B$ necessary for the evolution equations can be obtained from derivatives of the thermal mass and knowledge of the non-equilibrium microscopic parameters which enter the $\tilde{\cal H}_{3B}$ function.  However, in order to obtain the total energy density or pressures one needs to know $B$ itself.  Our procedure will be to integrate the dynamical equations to a very late proper time $\tau_f$ when the system is close to equilibrium and then integrate Eq.~(\ref{eq:B}) backwards in time from $\tau_f$ to the intial time $\tau_0$ subject to the boundary condition that $B(\tau_f) = B_{\rm eq}(T(\tau_f))$.

Using Eqs.~(\ref{eq:E-P-trans}), (\ref{eq:I-i}), and (\ref{eq:Ieq}) one can expand (\ref{eq:1st-mom-u}), (\ref{eq:xx-trans}), and (\ref{eq:zz-trans}) to obtain
\ba
&& 4 \tilde{\cal H}_3 \partial_\tau\log\lambda+\tilde{\Omega}_m\partial_\tau\log \hat{m} +\tilde{\Omega}_L\partial_\tau\log\alpha_z
+\tilde{\Omega}_T\partial_\tau\log\alpha_x^2 +\frac{\partial_\tau B}{\lambda^4}+\frac{\tilde{\Omega}_L}{\tau} = 0 \, , \hspace{1cm} \label{eq:final1-thermal} \\
&& 4\partial_\tau\log\alpha_x+\partial_\tau\log\alpha_z+5\partial_\tau\log\lambda
+ \partial_\tau \log\!\left(\hat{m}^3 K_3(\hat{m})\right)
+\frac{1}{\tau}
\nonumber \\ && \hspace{8cm} 
= \frac{1}{\tau_{\rm eq}}\left[\frac{1}{\alpha_x^4\alpha_z}\Big(\frac{T}{\lambda}\Big)^2\frac{K_3(\hat{m}_{\rm eq})}{K_3(\hat{m})}-1\right] \,, 
\label{eq:final2xx-thermal} 
\\
&& 2\partial_\tau\log\alpha_x+3\partial_\tau\log\alpha_z+5\partial_\tau\log\lambda
+ \partial_\tau \log\!\left(\hat{m}^3 K_3(\hat{m})\right)
+\frac{3}{\tau} \nonumber \\ && \hspace{8cm} 
= \frac{1}{\tau_{\rm eq}}\left[\frac{1}{\alpha_x^2\alpha_z^3}\Big(\frac{T}{\lambda}\Big)^2\frac{K_3(\hat{m}_{\rm eq})}{K_3(\hat{m})}-1\right] \,,
\label{eq:final2zz-thermal}
\ea
where $\tilde{\Omega}_T$, $\tilde{\Omega}_L$, and $\tilde{\Omega}_m$ are defined in App.~\ref{subapp:h-functions-1}.

One can perform some algebra to change the matching condition (\ref{eq:matching}) into a differential equation which is more convenient to solve since we then only have to solve a system of coupled ordinary differential equations.  Taking a derivative of Eq.~(\ref{eq:matching}) with respect to $\tau$ and using Eq.~(\ref{eq:1st-mom-u}), one obtains
\be
4 \tilde{\cal H}_{3,\rm eq} \partial_\tau\log T +\tilde{\Omega}_{m,\rm eq}\partial_\tau\log \hat{m}_{\rm eq}
+\frac{\tilde{\Omega}_L}{\tau}\Big(\frac{\lambda}{T}\Big)^4+\frac{\partial_\tau B}{T^4}=0\,.
\label{eq:final-matching-thermal}
\ee
In all equations above one can use Eq.~(\ref{eq:teq}) for $\tau_{\rm eq}$.

\subsection{Standard equation of state}
\label{sec:massless}

We now present the details of our implementation of the ``standard EoS'' method.  In this case, one takes the particles to be massless, $m \rightarrow0$, and hence $B\rightarrow 0$. For the massless transversally-symmetric case, Eqs.~(\ref{eq:E-P-trans}) become
\ba
{\cal E} &=& \bar{\cal H}_3({\boldsymbol\alpha}) \, \lambda^4 \, ,\nonumber \\
{\cal P}_T &=& \bar{\cal H}_{3T}({\boldsymbol\alpha}) \, \lambda^4 \, ,\nonumber \\
{\cal P}_L &=& \bar{\cal H}_{3L}({\boldsymbol\alpha}) \, \lambda^4 \,,
\label{eq:massless1}
\ea
where all ${\cal H}$-functions are defined in App.~\ref{subapp:h-functions-2}.  As we can see from the above equations, there is a multiplicative factorization of the energy density and pressures into a function that only depends on the anisotropy parameters and a function that only depends on the scale $\lambda$.  For a massless conformal Boltzmann gas, one has ${\cal E}_{\rm eq}(T) = 24 \pi \tilde{N}T^4$ and ${\cal P}_{\rm eq}(T) = 8\pi \tilde{N}T^4$.  Using these relations, one can rewrite Eqs.~(\ref{eq:massless1}) in terms of the equilibrium thermodynamic functions 
\ba
 {\cal E}&=&\frac{{\cal E}_{\rm eq}(\lambda)}{2} \alpha_x^4 \bar{{\cal H}}_2\Big(\frac{\alpha_z}{\alpha_x}\Big) ,\nonumber \\
  {\cal P}_T&=&\frac{3{\cal P}_{\rm eq}(\lambda)}{4} \alpha_x^4 \bar{{\cal H}}_{2T}\Big(\frac{\alpha_z}{\alpha_x}\Big) ,\nonumber \\
 {\cal P}_L&=&\frac{3{\cal P}_{\rm eq}(\lambda)}{2} \alpha_x^4 \bar{{\cal H}}_{2L}\Big(\frac{\alpha_z}{\alpha_x}\Big) .
 \label{eq:EPb}
\ea
These formulas suggest that, in order to impose a realistic EoS, one only has to replace ${\cal E}_{\rm eq}(\lambda)$ and ${\cal P}_{\rm eq}(\lambda)$ by the results obtained from lattice QCD calculations.

In order to obtain the necessary dynamical equations, one has to take the limit $m\rightarrow0$ of the equations obtained from the moments of the Boltzmann equation and substitute ${\cal E}$ and ${\cal P}_{T,L}$ from Eq.~(\ref{eq:EPb}). 
For the first moment equation, starting from Eq.~(\ref{eq:1st-mom-u}) and using Eq.~(\ref{eq:EPb}) one obtains
\be
\partial_\tau \log {\cal E}_{\rm eq}(\lambda) +(1+\chi)\partial_\tau \log \alpha_z +(3-\chi)\partial_\tau \log \alpha_x
 = -\frac{1}{\tau}-\frac{3 P_{\rm eq}(\lambda)}{\tau {\cal E}_{\rm eq}(\lambda)}\chi\,,\label{eq:final1-massless}
\ee
with $\chi\equiv \bar{{\cal H}}_{2L}/\bar{{\cal H}}_2$.  Taking the limit $m\rightarrow 0$ and $B\rightarrow 0$ of the second-moment equations (\ref{eq:final2xx-thermal}) and (\ref{eq:final2zz-thermal}), one obtains
\ba
4\partial_\tau\log\alpha_x+\partial_\tau\log\alpha_z+5\partial_\tau\log\lambda+\frac{1}{\tau} = \frac{1}{\tau_{\rm eq}}\Big[\Big(\frac{T}{\lambda}\Big)^5\frac{1}{\alpha_x^4\alpha_z}-1\Big] ,
\\
2\partial_\tau\log\alpha_x+3\partial_\tau\log\alpha_z+5\partial_\tau\log\lambda+\frac{3}{\tau} = \frac{1}{\tau_{\rm eq}}\Big[\Big(\frac{T}{\lambda}\Big)^5\frac{1}{\alpha_x^2\alpha_z^3}-1\Big] .
\label{eq:final2-massless}
\ea
For the matching relation which gives $T$ in terms of the microscopic parameters, one can use ${\cal E}(\lambda)= {\cal E}_{\rm eq}(T)$ and Eq.~(\ref{eq:final1-massless}) to find
\be
\partial_\tau\log{\cal E}_{\rm eq}(T) = -\frac{1}{\tau}-\frac{3 P_{\rm eq}(\lambda)}{\tau {\cal E}_{\rm eq}(\lambda)}\chi \,.
\label{eq:final-matching-massless}
\ee
In all equations above one can use Eq.~(\ref{eq:teq0}) for $\tau_{\rm eq}$.

\section{Results}
\label{sec:results}

In this section, we present the results of numerically integrating the dynamical equations using the ``standard EoS'' and the ``quasiparticle EoS'' methods.  In both cases, we specialize to the 0+1d case.  We take the initial proper time to be $\tau_0 = 0.25$ fm/c and the final time to be $\tau_f = 500$ fm/c.  In all cases, the initial temperature is taken to be \mbox{$T_0 = 600$ MeV} which is appropriate for LHC heavy-ion collisions at $\sqrt{s_{\rm NN}} = 2.76$ TeV.  The final time used here is very long compared to the timescales relevant for heavy-ion collisions, but we are interested in the late-time approach to isotropic thermal equilibrium in both approaches.  Additionally, as mentioned previously, in order to determine $B(\tau)$, we solve the the differential equation (\ref{eq:B}) by evolving it backwards in proper time subject to a boundary condition that $B(\tau_f) = B_{\rm eq}(T(\tau_f))$ and, consequently, we should evolve the system to a late proper-time at which the system is close to isotropic thermal equilibrium.  

Before proceedings to our results, we need to define one quantity which has yet to be defined, namely the bulk correction to the pressure.  In viscous hydrodynamics, the energy-momentum tensor is expressed generally as
\be
T^{\mu\nu} ={\cal E}_{\rm eq}u^\mu u^\nu-({\cal P}_{\rm eq}+\Pi)\Delta^{\mu\nu}+\pi^{\mu\nu}\, ,
\ee
where ${\cal E}_{\rm eq} = {\cal E}_{\rm eq}(T)$ and ${\cal P}_{\rm eq} = {\cal P}_{\rm eq}(T)$ are the equilibrium energy density and pressure evaluated at the effective temperature.  In the definition above, $\pi^{\mu\nu}$ is the shear tensor and $\Pi$ is the (isotropic) bulk correction.  Since $\pi^{\mu\nu}$ is a traceless tensor, $\pi^\mu_\mu=0$, which is transverse to the fluid four-velocity, $u_\mu\pi^{\mu\nu}=0$, one finds that the bulk correction can be computed from
\be
\Pi=-\frac{1}{3}\Delta_{\mu\nu}T^{\mu\nu}-{\cal P}_{\rm eq}=\frac{1}{3}({\cal P}_L+2 {\cal P}_T)-{\cal P}_{\rm eq}\,.
\ee
For the case of a temperature-dependent mass, one can use Eqs.~(\ref{eq:Peq}) and (\ref{eq:E-P-trans}).  For the massless case, one can use Eqs.~(\ref{eq:P-func}) and (\ref{eq:EPb}).

\begin{figure}[t]
\hspace{-6mm}
\includegraphics[width=1\linewidth]{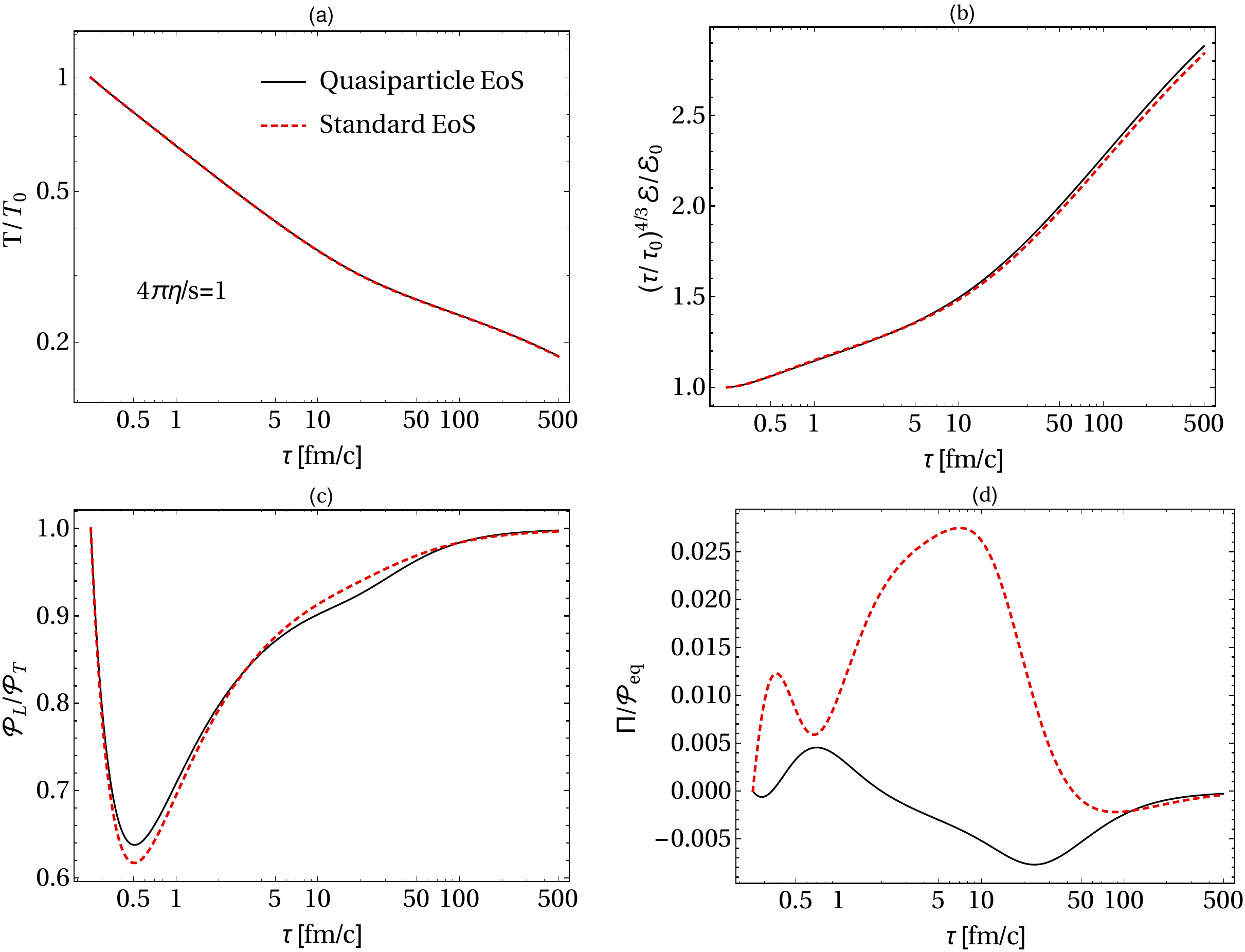}
\caption{The four panels show:  (a) the effective temperature scaled by $T_0$, (b) $(\tau/\tau_0)^{4/3}$ times the energy density scaled by the initial energy density, ${\cal E}_0$, (c) the pressure anisotropy, and (d) the bulk correction to the pressure scaled by ${\cal P}_{\rm eq}$.  For this figure we took $4 \pi \eta/s = 1$.}
\label{fig:plot-1-iso}
\end{figure}

\begin{figure}[t]
\hspace{-6mm}
\includegraphics[width=1\linewidth]{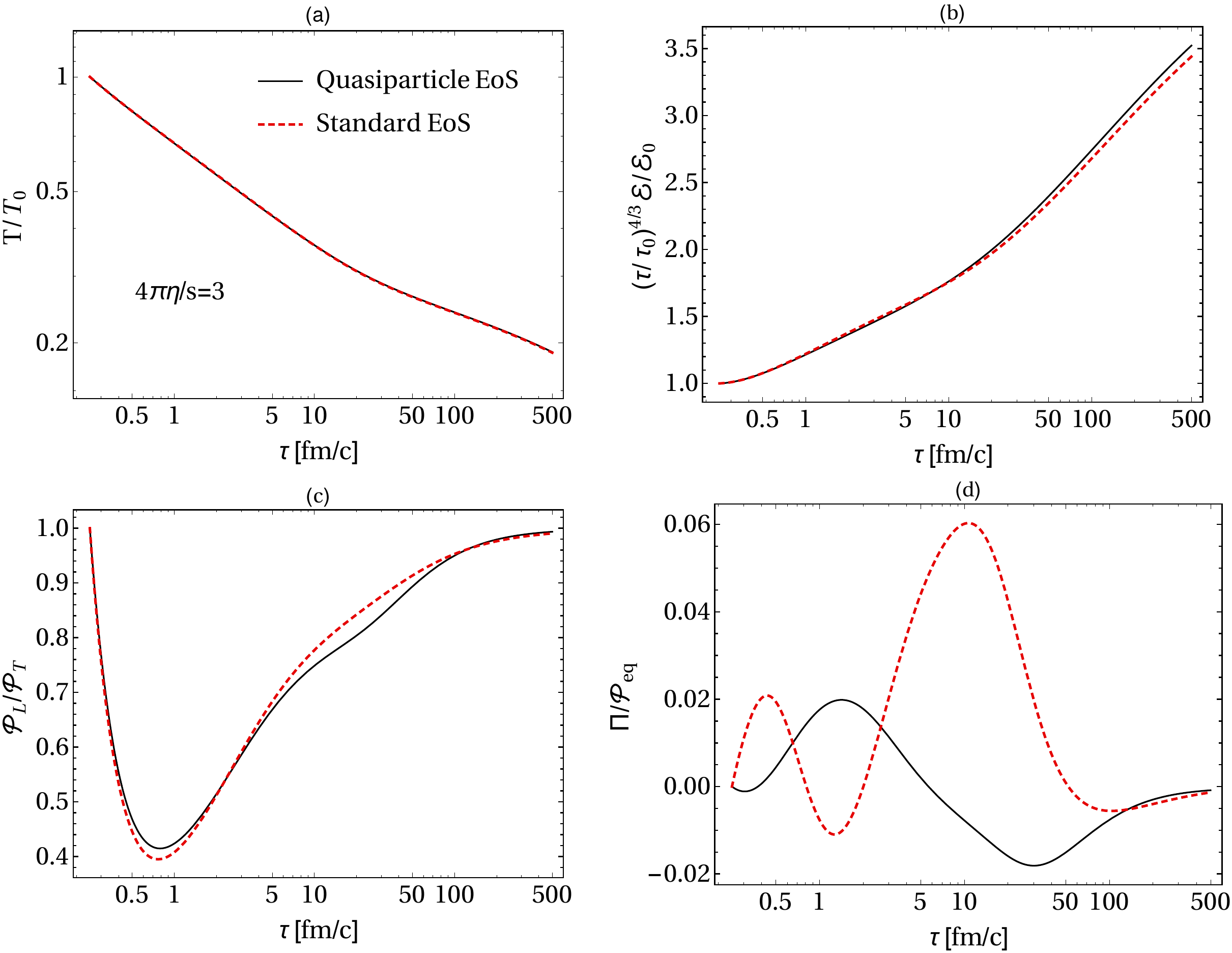}
\caption{Same as Fig.~\ref{fig:plot-1-iso} except here we take $4 \pi \eta/s = 3$.}
\label{fig:plot-3-iso}
\end{figure}

\begin{figure}[t]
\hspace{-6mm}
\includegraphics[width=1\linewidth]{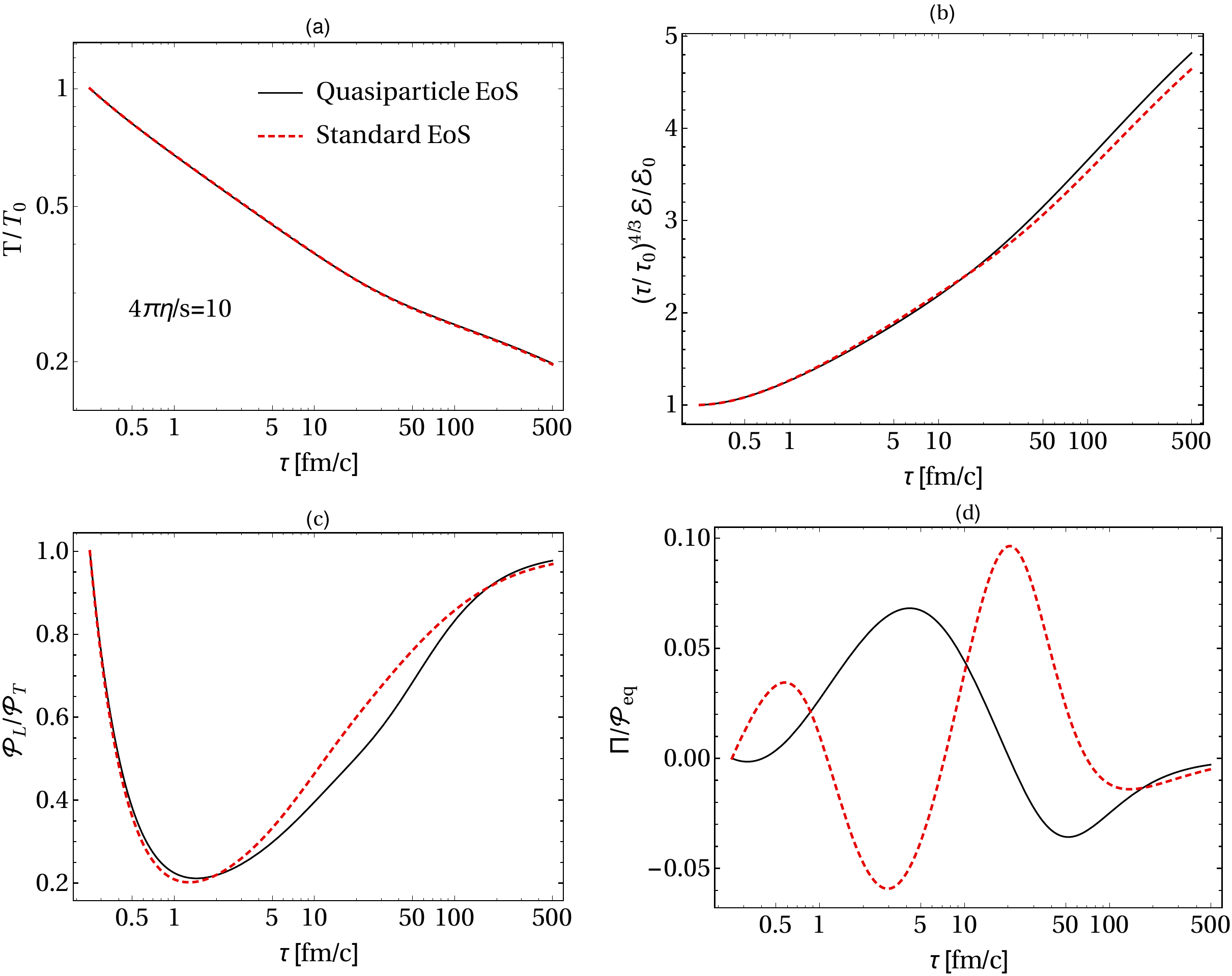}
\caption{Same as Fig.~\ref{fig:plot-1-iso} except here we take $4 \pi \eta/s = 10$.}
\label{fig:plot-10-iso}
\end{figure}

\subsection*{Numerical results}
\label{subsec:results}

We now turn to our numerical results.  In all plots, we compare the two methods for implementing the EoS in anisotropic hydrodynamics.  For the curves labeled ``Quasiparticle EoS'', we solve the dynamical equations specified in Sec.~\ref{sec:thermal-mass} and for the ``Standard EoS'', we solve those in Sec.~\ref{sec:massless}.  For purposes of the comparison, we match physical quantities rather than the microscopic parameters at $\tau_0$.  In practice, this means that we specify an initial temperature $T_0$, an initial momentum-space anisotropy quantified by ${\cal P}_{L,0}/{\cal P}_{T,0}$, and an initial bulk correction, $\Pi_0$.  In all results figures, we present four panels which correspond to: (a) the effective temperature scaled by $T_0$, (b) $(\tau/\tau_0)^{4/3}$ times the energy density scaled by ${\cal E}_0$, (c) the LRF pressure anisotropy, and (d) the bulk correction scaled by the equilibrium pressure, $\Pi/{\cal P}_{\rm eq}$.

In Figs.~\ref{fig:plot-1-iso} - \ref{fig:plot-10-iso} we present our results for the case of isotropic initial conditions.  In all panels, the microscopic parameters were adjusted to achieve ${\cal P}_{L,0}/{\cal P}_{T,0}=1$ and $\Pi_0=0$.  From panel (a) of this set of figures, we see that there is excellent agreement between the effective temperature predicted by each method for implementing the EoS.  In practice, we found that, for all initial conditions we considered, the maximum difference between the effective temperature obtained using the two approaches was less than on the order of 1\%.  To further explore the differences in the ``first order'' quantities, in panel (b) we have multiplied the scaled energy density by a factor of $(\tau/\tau_0)^{4/3}$.  If the system behaved as an ideal gas undergoing boost-invariant expansion in ideal hydrodynamics, then at late times this quantity should approach unity.  Any late-time deviations from unity are indicative of the corrections to ideal Bjorken scaling.  As we can see from panel (b) of Figs.~\ref{fig:plot-1-iso} - \ref{fig:plot-10-iso}, the energy density evolution obtained using the two approaches is quite close, with the largest difference between the two approaches being approximately 4\%.

Considering panel (c) of Figs.~\ref{fig:plot-1-iso} - \ref{fig:plot-10-iso}, we see that there are larger differences in the pressure anisotropy predicted by the two approaches.  For this quantity, we see differences as large as 20\%, however, the behavior of the pressure anisotropy is qualitatively the same overall.  Finally, we turn to panel (d) of Figs.~\ref{fig:plot-1-iso} - \ref{fig:plot-10-iso} which shows the bulk correction scaled by the equilibrium pressure.  As we see from these panels, there is a qualitative difference in the temporal evolution of the bulk correction when comparing the two approaches.  At late times, however, both approaches seem to converge to the same asymptotic limiting behavior.  Note that the differences in the pressure anisotropy and bulk correction are already self-consistently taken into account in the evolution of the temperature/energy density.  In this sense, despite having differences in the viscous corrections, the first order quantities seem to be quite insensitive to whether one uses the quasiparticle EoS method or the standard EoS method.  That being said, the differences seen in panels (c) and (d) could manifest themselves as differences in the particle spectra computed along the hypersurface if these two methods are applied to QGP phenomenology. 

\begin{figure}[t]
\hspace{-6mm}
\includegraphics[width=1\linewidth]{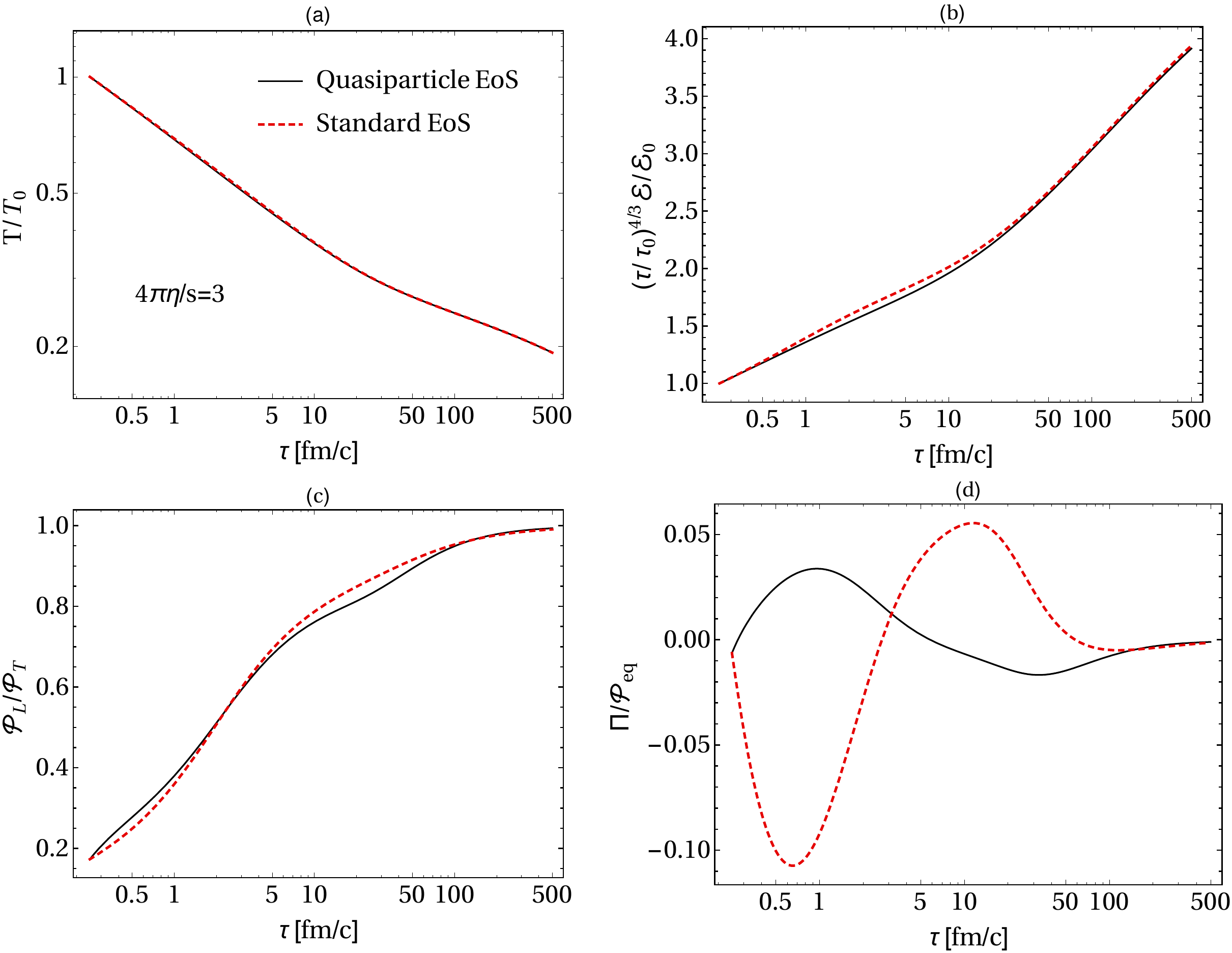}
\caption{Same as Fig.~\ref{fig:plot-3-iso} except here we take anisotropic initial conditions.}
\label{fig:plot-3-aniso}
\end{figure}

Finally, in Fig.~\ref{fig:plot-3-aniso} we present the same four panels, but in the case of an anisotropic initial condition with an oblate momentum-space anisotropy.  As can be seen from Fig.~\ref{fig:plot-3-aniso}, even for anisotropic initial conditions, the two EoS methods agree extremely well for the evolution the effective temperature and energy density.  However, similar to the case of isotropic initial conditions, we see somewhat larger differences in the evolution of the pressure anisotropy and qualitative differences in the evolution of the bulk pressure correction.  For both quantities, we see that the two methods have the same late-time asymptotic behavior.  Finally, we note that the behavior seen in Fig.~\ref{fig:plot-3-aniso} is indicative of the results we obtained for a variety of different non-equilibrium initial conditions.

\section{Conclusions and Outlook}
\label{sec:conclusions}

In this paper, we presented a new method for imposing a realistic EoS in the context of anisotropic hydrodynamics.  The method relies on a quasiparticle picture of the QGP, which is conceptually consistent with the kinetic theory method used to derive the required hydrodynamic evolution equations from the Boltzmann equation.  We discussed the fact that the introduction of a temperature-dependent quasiparticle mass requires an additional background contribution to the energy-momentum tensor.  We showed that requiring energy-momentum conservation results in a constraint equation on the background contribution which reduces to the constraint necessary to enforce thermodynamic consistency in the isotropic equilibrium limit as found by previous authors~\cite{Jeon:1995zm,Romatschke:2011qp}.  When solving the resulting dynamical equations, we allowed the background contribution to be a non-equilibrium quantity.  This was necessary to self-consistently implement the constraint equation.

By numerically solving the resulting dynamical equations in the 0+1d, we compared the results obtained using the quasiparticle EoS method with those obtained using the standard method for imposing a realistic EoS in anisotropic hydrodynamics.  We found that the temperature evolution obtained using the two methods was nearly identical and that there were only small differences in the pressure anisotropy.  However, we found that there were large qualitative differences in the evolution of the bulk pressure correction.  These conclusions were supported by the presentation of results for both isotropic and anisotropic initial conditions and also for different values of the shear viscosity to entropy density ratio, however, we internally checked a much larger set of initial conditions/parameter sets and found that these conclusions were generic.  We note, however, that the difference in the bulk pressure correction found does not necessarily imply large corrections for heavy-ion phenomenology.  As we have shown, first order quantities like the energy density are the same to within a few percent when comparing the two approaches.  That being said, the differences in the bulk pressure in particular could be important when fixing the form of the distribution function on the freezeout hypersurface.  

Looking forward, in a future work we will present numerical comparisons of the two approaches beyond the simple case of 0+1d expansion considered herein.  Additionally, it would be quite interesting to apply the quasiparticle EoS method to obtain the dynamical evolution for a non-conformal system with temperature-dependent masses within the context of second-order viscous hydrodynamics.  Finally, we note that it may be possible to construct exact solutions of the RTA Boltzmann equation for a system of particles with temperature-dependent masses using methods similar to those in Refs.~\cite{Florkowski:2013lza,Florkowski:2013lya,Florkowski:2014sfa}.  Additionally, for the case of quasiparticle masses that are linear in the temperature, it may be possible to exactly solve the RTA Boltzmann equation subject to Gubser flow similar to Refs.~\cite{Denicol:2014xca,Denicol:2014tha}.

\acknowledgments{ We thank W. Florkowski, P. Romatschke, and R. Ryblewski for useful conversations.  M. Strickland and M. Nopoush were supported by the U.S. Department of Energy, Office of Science, Office of Nuclear Physics under Awards No.~DE-SC0013470.  M. Alqahtani was supported by a PhD fellowship from the University of Dammam.}

\appendix 

\section{Explicit formulas for derivatives}
\label{app:identities}

In this section, first we introduce the notations used in derivation of the general moment-based hydrodynamics equations and then, by taking the appropriate limits, we simplify them for the transversally-homogeneous 0+1d case. Using the definitions
\ba
{\cal D}&\equiv&\cosh(\vartheta-\varsigma)\partial_\tau+\frac{1}{\tau}\sinh(\vartheta-\varsigma)\partial_\varsigma\nonumber\,, \\
\tilde{{\cal D}}&\equiv&\sinh(\vartheta-\varsigma)\partial_\tau+\frac{1}{\tau}\cosh(\vartheta-\varsigma)\partial_\varsigma\,, \\
\nabla_\perp\cdot{\bf u}_\perp &\equiv&\partial_x u_x+\partial_y u_y \nonumber\,, \\
{\bf u}_\perp\cdot\nabla_\perp &\equiv& u_x\partial_x+u_y\partial_y \nonumber\,,\\
{\bf u}_\perp\times\nabla_\perp &\equiv& u_x \partial_y-u_y\partial_x\,,
\label{eq:identities-gen}
\ea
and four-vectors defined in Eq.~(\ref{eq:4vectors}) one has
\ba
D_u&\equiv&u^\mu \partial_\mu=u_0{\cal D}+{\bf u}_\perp\cdot\nabla_\perp \nonumber\,,\\
D_x&\equiv&X^\mu \partial_\mu=u_\perp{\cal D}+\frac{u_0}{u_\perp}({\bf u}_\perp\cdot\nabla_\perp)\nonumber\,,\\
D_y&\equiv&Y^\mu \partial_\mu=\frac{1}{u_\perp}({\bf u}_\perp\times\nabla_\perp)\nonumber\,,\\
D_z&\equiv&Z^\mu \partial_\mu=\tilde{{\cal D}}\,.
\ea
The divergences are defined as
\ba 
\theta_u&\equiv&\partial_\mu u^\mu={\cal D}u_0+u_0\tilde{{\cal D}}\vartheta+\nabla_\perp\cdot{\bf u}_\perp\nonumber\,,\\
\theta_x&\equiv&\partial_\mu X^\mu={\cal D}u_\perp+u_\perp\tilde{{\cal D}}\vartheta+\frac{u_0}{u_\perp}(\nabla_\perp\cdot{\bf u}_\perp)-\frac{1}{u_0 u_\perp^2}({\bf u}_\perp\cdot\nabla_\perp) u_\perp\nonumber\,,\\
\theta_y&\equiv&\partial_\mu Y^\mu=-\frac{1}{u_\perp}({\bf u}_\perp\cdot\nabla_\perp) \varphi\nonumber\,, \\
\theta_z&\equiv&\partial_\mu Z^\mu={\cal D}\vartheta\,,
\label{eq:deriv-gen}
\ea
where $\varphi=\tan^{-1}(u_y/u_x)$.
\ba
u_\mu D_\alpha X^\mu &=&\frac{1}{u_0}D_\alpha u_\perp \nonumber\,, \\
u_\mu D_\alpha Y^\mu &=& u_\perp D_\alpha\varphi \nonumber \,, \\
u_\mu D_\alpha Z^\mu &=& u_0 D_\alpha\vartheta \nonumber \,,\\
X_\mu D_\alpha Y^\mu &=& u_0 D_\alpha\varphi \nonumber \,,\\
X_\mu D_\alpha Z^\mu &=& u_\perp D_\alpha\vartheta \nonumber\,, \\
Y_\mu D_\alpha Z^\mu &=& 0\,, \label{eq:iden2-gen}
\ea
where $\alpha\in\{u,x,y,z\}$.  Note that contractions such as $X^\mu D_\alpha u_\mu$ are also non-vanishing, however, such terms can be written in terms of the expressions above by using the orthogonality of the basis vectors, i.e. $D_\alpha(X^\mu u_\mu) = 0$ implies that $X^\mu D_\alpha u_\mu = -u_\mu D_\alpha X^\mu$.

\subsection{Simplification for 1+1d}
In the case of boost-invariant and cylindrically-symmetric flow one has $\varphi \rightarrow \phi$ and
$\vartheta \rightarrow \varsigma$, where $\varsigma$ is the spatial rapidity. Using $u_\perp\equiv\sinh\theta_\perp$, one can rewrite (\ref{eq:identities-gen}) as
\ba
{\cal D}&=&\partial_\tau\nonumber\,,\\
\tilde{{\cal D}}&=&\frac{1}{\tau}\partial_\varsigma\,, \\
\nabla_\perp\cdot{\bf u}_\perp &=& \partial_r u_\perp+\frac{u_\perp}{r} \nonumber\,, \\
{\bf u}_\perp\cdot\nabla_\perp&=& u_\perp\partial_r \nonumber\,,\\
{\bf u}_\perp\times\nabla_\perp &=&\frac{u_\perp}{r}\partial_\phi \nonumber\,.
\ea
Also, the identities in (\ref{eq:deriv-gen}) become
\ba
D_u&=&\cosh\theta_\perp \partial_\tau+\sinh\theta_\perp\partial_r\,,\\
D_x&=&\sinh\theta_\perp\partial_\tau+\cosh\theta_\perp\partial_r\,,\\
D_y&=&\frac{1}{r}\partial_\phi\,, \\
D_z&=&\frac{1}{\tau}\partial_\varsigma\,,\\
\theta_u&=&\cosh\theta_\perp\Big(\frac{1}{\tau}+\partial_r\theta_\perp\Big)+\sinh\theta_\perp\Big(\frac{1}{r}+\partial_\tau\theta_\perp\Big) ,\\
\theta_x&=&\sinh\theta_\perp\Big(\frac{1}{\tau}+\partial_r\theta_\perp\Big)+\cosh\theta_\perp\Big(\frac{1}{r}+\partial_\tau\theta_\perp\Big) ,\\
\theta_y&=&\theta_z=0\,.
\ea
In this limit, the only non-vanishing terms in (\ref{eq:iden2-gen}) are
\ba 
u_\mu D_u X^\mu &=& D_u\theta_\perp \nonumber\,, \\
u_\mu D_x X^\mu &=&D_x\theta_\perp \nonumber \,,\\
u_\mu D_y Y^\mu&=&\frac{1}{r}\sinh\theta_\perp \nonumber\,, \\
u_\mu D_z Z^\mu&=&\frac{1}{\tau}\cosh\theta_\perp \nonumber\,, \\
X_\mu D_y Y^\mu&=&\frac{1}{r}\cosh\theta_\perp \nonumber\,, \\
X_\mu D_z Z^\mu&=&\frac{1}{\tau} \sinh\theta_\perp\,.
\ea 
%
\subsection{Simplification for 0+1d}
For this case, one has $\theta_\perp=0$ and
\ba
D_u&=&\partial_\tau\,,\\
D_x&=&\partial_r\,,\\
D_y&=&\frac{\partial_\phi}{r}\,, \\
D_z&=&\frac{\partial_\varsigma}{\tau}\,,\\
\theta_u&=&\frac{1}{\tau}\,,\\
\theta_x&=&\frac{1}{r}\,,\\
\theta_y&=&\theta_z=0\,.
\ea
In this limit, the only non-vanishing terms in (\ref{eq:iden2-gen}) are
\ba 
u_\mu D_z Z^\mu&=&\frac{1}{\tau} \nonumber\,, \\
X_\mu D_y Y^\mu&=&\frac{1}{r} \nonumber \,.
\ea 

\section{special functions}
\label{app:h-functions}

In this section, we provide definitions of the special functions appearing in the body of the text. We start by introducing 
\ba
 {\cal H}_2(y,z) &\equiv&
y \int_{-1}^1 dx  \; 
\sqrt{(y^2-1)x^2 + z^2+1}
\nonumber \\
&=& \frac{y}{\sqrt{y^2-1}} \left[ (z^2+1)
\tanh^{-1} \sqrt{\frac{y^2-1}{y^2+z^2}} + \sqrt{(y^2-1)(y^2+z^2)} \, \right] ,
\label{eq:H2}
\ea
\ba
{\cal H}_{2T}(y,z) &\equiv&
 y \, 
\int\limits_{-1}^1 \frac{dx (1-x^2) }{
\, \sqrt{(y^2-1)x^2 + z^2+1}} 
\label{eq:H2T} \nonumber \\
&=& \frac{y}{(y^2-1)^{3/2}}
\left[\left(z^2+2y^2-1\right) 
\tanh^{-1}\sqrt{\frac{y^2-1}{y^2+z^2}}
-\sqrt{(y^2-1)(y^2+z^2)} \right] , \hspace{1cm}
\ea
\ba
{\cal H}_{2L}(y,z) &\equiv& y^3 \, 
 \int\limits_{-1}^1 \frac{ dx\,x^2 }{\, \sqrt{(y^2-1)x^2 + z^2+1}}
 \nonumber \\
&=& \frac{y^3}{(y^2-1)^{3/2}}
\left[
\sqrt{(y^2-1)(y^2+z^2)}-(z^2+1)
\tanh^{-1}\sqrt{\frac{y^2-1}{y^2+z^2}} \,\,\right]. 
\label{eq:H2L}
\ea
Derivatives of these functions satisfy the following relations
\ba
\frac{\partial {\cal H}_2(y,z)}{\partial y}&=&\frac{1}{y}\Big[{\cal H}_2(y,z)+{\cal H}_{2L}(y,z)\Big] , \\
\frac{\partial {\cal H}_2(y,z)}{\partial z}&=&\frac{1}{z}\Big[{\cal H}_2(y,z)-{\cal H}_{2L}(y,z)-{\cal H}_{2T}(y,z)\Big] .
\ea
 
\subsection{Massive Case} 
\label{subapp:h-functions-1}

The ${\cal H}$-functions appearing in the definitions of components of the energy-momentum tensor are 
\ba
{\cal H}_3({\boldsymbol\alpha},\hat{m}) &\equiv &  \tilde{N} \alpha_x \alpha_y
\int_0^{2\pi} d\phi \, \alpha_\perp^2 \int_0^\infty d\hat{p} \, \hat{p}^3  f_{\rm eq}\!\left(\!\sqrt{\hat{p}^2 + \hat{m}^2}\right) {\cal H}_2\!\left(\frac{\alpha_z}{\alpha_\perp},\frac{\hat{m}}{\alpha_\perp \hat{p}} \right),
\label{eq:h3gen}
\\
{\cal H}_{3x}({\boldsymbol\alpha},\hat{m}) &\equiv &  \tilde{N}\alpha_x^3 \alpha_y
\int_0^{2\pi} d\phi \, \cos^2\phi \int_0^\infty d\hat{p} \, \hat{p}^3  f_{\rm eq}\!\left(\!\sqrt{\hat{p}^2 + \hat{m}^2}\right) {\cal H}_{2T}\!\left(\frac{\alpha_z}{\alpha_\perp},\frac{\hat{m}}{\alpha_\perp \hat{p}} \right),
\;\;\;\;
\label{eq:h3xgen}
\\
{\cal H}_{3y}({\boldsymbol\alpha},\hat{m}) &\equiv &  \tilde{N}\alpha_x \alpha_y^3
\int_0^{2\pi} d\phi \, \sin^2\phi \int_0^\infty d\hat{p} \, \hat{p}^3  f_{\rm eq}\!\left(\!\sqrt{\hat{p}^2 + \hat{m}^2}\right) {\cal H}_{2T}\!\left(\frac{\alpha_z}{\alpha_\perp},\frac{\hat{m}}{\alpha_\perp \hat{p}} \right) ,
\label{eq:h3ygen}
\\
{\cal H}_{3T}({\boldsymbol\alpha},\hat{m}) &\equiv &  \frac{1}{2} \Big[ {\cal H}_{3x}({\boldsymbol\alpha},\hat{m}) + {\cal H}_{3y}({\boldsymbol\alpha},\hat{m}) \Big] ,
\label{eq:h3tgen}
\\
{\cal H}_{3L}({\boldsymbol\alpha},\hat{m}) &\equiv &  \tilde{N} \alpha_x \alpha_y
\int_0^{2\pi} d\phi \, \alpha_\perp^2 \int_0^\infty d\hat{p} \, \hat{p}^3  f_{\rm eq}\!\left(\!\sqrt{\hat{p}^2 + \hat{m}^2}\right) {\cal H}_{2L}\!\left(\frac{\alpha_z}{\alpha_\perp},\frac{\hat{m}}{\alpha_\perp \hat{p}} \right) ,
\label{eq:h3lgen} 
\\
{\cal H}_{3m}({\boldsymbol\alpha},\hat{m}) &\equiv &  \tilde{N}\alpha_x\alpha_y\hat{m}^2
\int_0^{2\pi} d\phi \, \alpha_\perp^2\int_0^\infty d\hat{p} \,\hat{p}^3 \frac{f_{\rm eq}\!\left(\!\sqrt{\hat{p}^2 + \hat{m}^2}\right)}{\sqrt{\hat{p}^2+\hat{m}^2}}   {\cal H}_2\!\left(\frac{\alpha_z}{\alpha_\perp},\frac{\hat{m}}{\alpha_\perp \hat{p}} \right) ,
\label{eq:h3mgen} 
\\
{\cal H}_{3B}({\boldsymbol\alpha},\hat{m}) &\equiv &  \tilde{N}\alpha_x\alpha_y
\int_0^{2\pi} d\phi\int_0^\infty d\hat{p} \, \hat{p} f_{\rm eq}\!\left(\!\sqrt{\hat{p}^2 + \hat{m}^2}\right) {\cal H}_{\rm 2B}\!\left(\frac{\alpha_z}{\alpha_\perp},\frac{\hat{m}}{\alpha_\perp \hat{p}} \right) ,
\label{eq:h3Bgen} 
\\
\Omega_T({\boldsymbol\alpha},\hat{m}) &\equiv& {\cal H}_3+{\cal H}_{3T}\,,  \\
\Omega_L({\boldsymbol\alpha},\hat{m}) &\equiv& {\cal H}_3+{\cal H}_{3L}\,,  \\
\Omega_m({\boldsymbol\alpha},\hat{m}) &\equiv& {\cal H}_3-{\cal H}_{3L}-2{\cal H}_{3T}-{\cal H}_{3m}\,,
\ea
where $\alpha_\perp^2 \equiv \alpha_x^2 \cos^2\phi  + \alpha_y^2 \sin^2\phi $ and
\be
{\cal H}_{2B}(y,z)\equiv {\cal H}_{2T}(y,z)+ \frac{{\cal H}_{2L}(y,z)}{y^2}=\frac{2}{\sqrt{y^2-1}}\tanh^{-1} \sqrt{\frac{y^2-1}{y^2+z^2}} \, .
\ee
For a 0+1d system one has $\alpha_x = \alpha_y$ such that $\alpha_\perp = \alpha_x$ and $\tilde{\cal H}_{3T} \equiv  \tilde{\cal H}_{3x}  = \tilde{\cal H}_{3y}$, so that one obtains
\ba
\tilde{\cal H}_3({\boldsymbol\alpha},\hat{m}) &\equiv&  2 \pi \tilde{N} \alpha_x^4
\int_0^\infty d\hat{p} \, \hat{p}^3  f_{\rm eq}\!\left(\!\sqrt{\hat{p}^2 + \hat{m}^2}\right) {\cal H}_2\!\left(\frac{\alpha_z}{\alpha_x},\frac{\hat{m}}{\alpha_x\hat{p}} \right) ,
\label{eq:h3tilde}
\\
\tilde{\cal H}_{3T}({\boldsymbol\alpha},\hat{m}) &\equiv&  \pi \tilde{N} \alpha_x^4
\int_0^\infty d\hat{p} \, \hat{p}^3  f_{\rm eq}\!\left(\!\sqrt{\hat{p}^2 + \hat{m}^2}\right) {\cal H}_{2T}\!\left(\frac{\alpha_z}{\alpha_x},\frac{\hat{m}}{\alpha_x\hat{p}} \right) ,
\label{eq:h3ttilde}
\\
\tilde{\cal H}_{3L}({\boldsymbol\alpha},\hat{m}) &\equiv&  2 \pi \tilde{N} \alpha_x^4
\int_0^\infty d\hat{p} \, \hat{p}^3  f_{\rm eq}\!\left(\!\sqrt{\hat{p}^2 + \hat{m}^2}\right) {\cal H}_{2L}\!\left(\frac{\alpha_z}{\alpha_x},\frac{\hat{m}}{\alpha_x\hat{p}} \right) ,
\label{eq:h3ltilde}
\\
\tilde{\cal H}_{3m}({\boldsymbol\alpha},\hat{m}) &\equiv&  2 \pi \tilde{N} \alpha_x^4 \hat{m}^2
\int_0^\infty d\hat{p} \, \hat{p}^3  \frac{f_{\rm eq}\!\left(\!\sqrt{\hat{p}^2 + \hat{m}^2}\right)}{\sqrt{\hat{p}^2 + \hat{m}^2}} {\cal H}_2\!\left(\frac{\alpha_z}{\alpha_x},\frac{\hat{m}}{\alpha_x\hat{p}} \right) ,
\label{eq:h3mtilde}
\\
\tilde{\cal H}_{3B}({\boldsymbol\alpha},\hat{m}) &\equiv &  2\pi \tilde{N}\alpha_x^2
\int_0^\infty d\hat{p} \, \hat{p} f_{\rm eq}\!\left(\!\sqrt{\hat{p}^2 + \hat{m}^2}\right) {\cal H}_{\rm 2B}\!\left(\frac{\alpha_z}{\alpha_x},\frac{\hat{m}}{\alpha_x \hat{p}} \right) .
\label{eq:h3Btilde} 
\ea
Also, derivatives of $\tilde{\cal H}_3$ satisfy
\ba
\frac{\partial \tilde{\cal H}_3}{\partial\alpha_x}&=&\frac{2}{\alpha_x}\tilde\Omega_T\,, \\
\frac{\partial \tilde{\cal H}_3}{\partial\alpha_z}&=&\frac{1}{\alpha_z}\tilde\Omega_L\,, \\
\frac{\partial \tilde{\cal H}_3}{\partial\hat{m}}&=&\frac{1}{\hat{m}}\tilde\Omega_m\,.
\ea
For the isotropic equilibrium case, one has $\alpha_i\rightarrow 1$, $\lambda\rightarrow T$, and $\hat{m}\rightarrow\hat{m}_{\rm eq}$
\ba
\tilde{\cal H}_{3,\rm eq}(\hat{m}_{\rm eq}) &=&  4 \pi \tilde{N} \hat{m}^2_{\rm eq}\Big[\hat{m}_{\rm eq}K_1(\hat{m}_{\rm eq})+3K_2(\hat{m}_{\rm eq})\Big] \,,
\label{eq:h3iso}
\\
\tilde{\cal H}_{3T,\rm eq}(\hat{m}_{\rm eq}) &=& \tilde{\cal H}_{3L,\rm eq}(\hat{m}_{\rm eq}) =4 \pi \tilde{N} \hat{m}^2_{\rm eq}K_2(\hat{m}_{\rm eq}) \,,
\label{eq:h3tiso}
\\
\tilde{\cal H}_{3m,\rm eq}(\hat{m}_{\rm eq}) &=& 4 \pi \tilde{N} \hat{m}^4_{\rm eq}K_2(\hat{m}_{\rm eq}) \,.
\label{eq:h3miso}
\ea
%

\subsection{Massless Case}
\label{subapp:h-functions-2}

Taking the massless limit of Eqs.~(\ref{eq:h3gen}) - (\ref{eq:h3Bgen}) one obtains
\ba
\hat{{\cal H}}_3 ({\boldsymbol\alpha})&\equiv &\lim_{m\rightarrow 0}{\cal H}_3({\boldsymbol\alpha},\hat{m}) =  6\tilde{N} \alpha_x \alpha_y
\int_0^{2\pi} d\phi \, \alpha_\perp^2 \bar{{\cal H}}_2\Big(\frac{\alpha_z}{\alpha_\perp} \Big) ,
\label{eq:h30}
\\
\hat{{\cal H}}_{3x} ({\boldsymbol\alpha})&\equiv &\lim_{m\rightarrow 0}{\cal H}_{3x}({\boldsymbol\alpha},\hat{m}) =  6 \tilde{N}\alpha_x^3 \alpha_y
\int_0^{2\pi} d\phi \, \cos^2\phi \, \bar{{\cal H}}_{2T}\Big(\frac{\alpha_z}{\alpha_\perp} \Big) ,
\label{eq:h3x0}
\\
\hat{{\cal H}}_{3y} ({\boldsymbol\alpha})&\equiv &\lim_{m\rightarrow 0}{\cal H}_{3y}({\boldsymbol\alpha},\hat{m}) =  6 \tilde{N}\alpha_x \alpha_y^3
\int_0^{2\pi} d\phi \, \sin^2\phi \, \bar{{\cal H}}_{2T}\Big(\frac{\alpha_z}{\alpha_\perp} \Big) ,
\label{eq:h3y0}
\\
\hat{{\cal H}}_{3L} ({\boldsymbol\alpha})&\equiv &\lim_{m\rightarrow 0}{\cal H}_{3L}({\boldsymbol\alpha},\hat{m}) =  6\tilde{N} \alpha_x \alpha_y
\int_0^{2\pi} d\phi \, \alpha_\perp^2 \bar{{\cal H}}_{2L}\Big(\frac{\alpha_z}{\alpha_\perp} \Big) ,
\label{eq:h3l0}
\\
\hat{{\cal H}}_{3m} ({\boldsymbol\alpha})&\equiv &\lim_{m\rightarrow 0}{\cal H}_{3m}({\boldsymbol\alpha},\hat{m})=0\,,
\label{eq:h3m0}
\ea
where $\bar{{\cal H}}_{2,2T,2L}(y) \equiv {\cal H}_{2,2T,2L}(y,0)$.
In the transversally-symmetric case, $\alpha_x=\alpha_y$ and $\bar{\cal H}_{3T} \equiv  \bar{\cal H}_{3x}  = \bar{\cal H}_{3y}$, and the functions above simplify to
\ba
\bar{{\cal H}}_3 ({\boldsymbol\alpha}) &=&  12\pi\tilde{N} \alpha_x^4
 \bar{{\cal H}}_2\Big(\frac{\alpha_z}{\alpha_x} \Big) ,
\label{eq:h30-trans}
\\
\bar{{\cal H}}_{3T} ({\boldsymbol\alpha})&=&  6\pi \tilde{N}\alpha_x^4  \bar{{\cal H}}_{2T}\Big(\frac{\alpha_z}{\alpha_x} \Big) ,
\label{eq:h3t0-trans}
\\
\bar{{\cal H}}_{3L} ({\boldsymbol\alpha}) &=&  12\pi\tilde{N} \alpha_x^4  \bar{{\cal H}}_{2L}\Big(\frac{\alpha_z}{\alpha_x} \Big) .
\label{eq:h3l0-trans}
\ea
In the isotropic equilibrium case, one has  $\alpha_i\rightarrow 1$ and $\lambda\rightarrow T$, and, as a result,
\ba
\bar{{\cal H}}_{3,{\rm eq}}  &=&  24\pi\tilde{N}\,,
\label{eq:h30-trans-iso}
\\
\bar{{\cal H}}_{3T,{\rm eq}} &=&\bar{{\cal H}}_{3L,{\rm eq}}=  8\pi \tilde{N} \, .
\label{eq:h3t0-trans-iso}
\ea

\bibliography{thermal-mass}

\end{document}